\documentclass[onecolumn,preprintnumbers,amsmath,amssymb,5pt]{revtex4}

\usepackage{graphicx}
\usepackage{dcolumn}
\usepackage{bm}
\usepackage{multirow}
\usepackage{tabularx}
\usepackage{indentfirst}
\usepackage{color}
\usepackage{chemarrow}
\usepackage{nomencl}
\usepackage[normalem]{ulem}
\makenomenclature

\usepackage{epstopdf}

\begin{document}

\title{Statistical Mechanics and Kinetics of Amyloid Fibrillation}
\author{Liu Hong$^{1,*}$,\;\; Chiu Fan Lee$^{2,*}$,\;\;Ya Jing Huang$^{1}$ \\
\footnotesize $^1$Zhou Pei-Yuan Center for Applied Mathematics, Tsinghua University, Beijing 100084, P.R. China \\
\footnotesize $^2$Department of Bioengineering, Imperial College London, South Kensington Campus, London SW7 2AZ, UK\\
\footnotesize $^*$Correspondence: zcamhl@tsinghua.edu.cn, c.lee@imperial.ac.uk\\}

\begin{abstract}
Amyloid fibrillation is a protein self-assembly phenomenon that is intimately related to well-known human neurodegenerative diseases. During the past few decades, striking advances have been achieved in our understanding of the physical origin of this phenomenon and they constitute the contents of this review. Starting from a minimal model of amyloid fibrils, we explore systematically the equilibrium and kinetic aspects of amyloid fibrillation in both dilute and semi-dilute limits. We then incorporate further molecular mechanisms into the analyses. We also discuss the mathematical foundation of kinetic modeling based on chemical mass-action equations, the quantitative linkage with experimental measurements, as well as the procedure to perform global fitting.
\end{abstract}

\maketitle

\emph{L.H. and Y.J.H. would like to dedicate this paper to the memory of Prof. Chia-Chiao Lin (1916-2013), a great applied mathematician, a beloved advisor and a dear friend, on his 100-year anniversary.}

\nomenclature[01]{$[A_i]$}{Concentration of filaments of size $i$}
\nomenclature[02]{$[C_i]$}{Concentration of cells in state $i$}
\nomenclature[03]{$N_s(n_s)$}{Number (concentration) of s-mer}
\nomenclature[04]{$N_s^{(\beta)}(n_s^{(\beta)})$}{Number (concentration) of monomers in the beta-sheet configuration}
\nomenclature[05]{$N_s^R(n_s^R)$}{Number (concentration) of monomers in the random coil configuration}
\nomenclature[11]{$m$}{Monomer concentration}
\nomenclature[12]{$P$}{Number concentration of total aggregates}
\nomenclature[13]{$M$}{Mass concentration of total aggregates}
\nomenclature[14]{$P_{oli}$}{Number concentration of oligomers}
\nomenclature[15]{$M_{oli}$}{Mass concentration of oligomers}
\nomenclature[16]{$m_{tot}$}{Total concentration of proteins}
\nomenclature[17]{$m_0$}{Initial monomer concentration}
\nomenclature[18]{$P_0$}{Initial number concentration of aggregates}
\nomenclature[19]{$M_0$}{Initial mass concentration of aggregates}
\nomenclature[31]{$n_c$}{Critical nucleus size for primary nucleation}
\nomenclature[32]{$n_2$}{Critical nucleus size for secondary nucleation}
\nomenclature[33]{$K_m$}{Critical saturation concentration for elongation}
\nomenclature[34]{$K_s$}{Critical saturation concentration for secondary nucleation}
\nomenclature[35]{$m^*_F$}{Critical fibrillar concentration}
\nomenclature[36]{$m^*_M$}{Critical micellar concentration}
\nomenclature[41]{$k_e^+$}{Rate constant for monomer association}
\nomenclature[42]{$k_e^-$}{Rate constant for monomer dissociation}
\nomenclature[43]{$k_n$}{Rate constant for primary nucleation}
\nomenclature[44]{$k_2$}{Rate constant for surface catalysed secondary nucleation}
\nomenclature[45]{$k_f^+$}{Rate constant for filaments fragmentation}
\nomenclature[46]{$k_f^-$}{Rate constant for filaments annealing}
\nomenclature[51]{$k_c^+$}{Forward reaction rate constant for conformation conversion}
\nomenclature[52]{$k_c^-$}{Backward reaction rate constant for conformation conversion}
\nomenclature[53]{$k_b^+$}{Rate constant for membrane binding}
\nomenclature[54]{$k_b^-$}{Rate constant for membrane unbinding}
\nomenclature[61]{$k_{app}$}{Apparent fiber growth rate}
\nomenclature[62]{$k_{max}$}{Maximal fiber growth rate}
\nomenclature[63]{$t_{1/2}$}{Half-time for fibrillation}
\nomenclature[64]{$t_{lag}$}{Lag-time for fibrillation}
\printnomenclature

\section{Introduction}
The importance of understanding  amyloid fibrillation comes not only from its intimate relation to amyloid diseases, such as the well-known Alzheimer's, Huntington's and Parkinson's diseases \cite{dobson2003protein, chiti2006protein}, but also from its physical simplicity and universality as a typical self-assembling phenomenon of linear biomolecules \cite{schreck2013statistical}. Various thermodynamic and kinetic approaches borrowed from classical polymer statistical mechanics, the kinetics of chemical reactions as well as non-equilibrium processes have been developed and applied to experimentally and biologically relevant amyloid systems with great success \cite{morris2009protein, cohen2012macroscopic, gillam2013modelling}. Related fruitful results, developments and applications in the past decades constitute the focus of our current paper: a self-contained review on the thermodynamic and kinetics of amyloid fibrillation.

Thermodynamics and kinetics are two sides of the same coin. The latter deals with time-dependent fibrillation processes in general; while the former is more focused on the final time-independent properties of the amyloid system -- the equilibrium state. In the current review, we will present the thermodynamics and kinetics of amyloid fibrillation separately in order to keep each part clear and self-contained. But readers should bear in mind of the intrinsic correlations between those two descriptions, like requirements on reaction rate constants for various fibrillation processes in order to guarantee the existence of a genuine thermodynamic equilibrium state \cite{hong2011lattice}.

The whole review is organized into three major sections. The first one is focused on the thermodynamics of amyloid fibrillation by using the language of statistical mechanics; the next two are devoted to kinetic descriptions based on chemical mass-action equations. To be specific, the former provides a systematic exploration of various amyloid fibrillation processes, including both model formulation and analysis; while the latter is about the mathematical foundation of kinetic modeling as well as its linkage with experimental observations.

\section{Statistical mechanics of amyloid fibrillation}
Although polymer physics is a well established field of science,
novel physics governing the behaviour of the systems is still being uncovered. In the case of self-assembling biopolymers, the novelty comes from the fact that the binding energy driving the polymerization process is relatively low compared to covalently bonded polymers. Hence, polymer breakage and re-joining can potentially contribute to the polymerization kinetics at an experimentally and physiologically relevant scale. The polymeric system is thus called ``living'' since every polymer can shrink through breakage and grow through elongation {\it via} monomer additions and through end-to-end joining with another polymer. Taking these processes into account are important for the complete description of the kinetics of self-assembling polymers. In this section, we will focus purely on how a system of living semi-flexible polymers behave at thermal equilibrium.

\begin{figure}[h]
\begin{center}
\includegraphics[scale=.6]{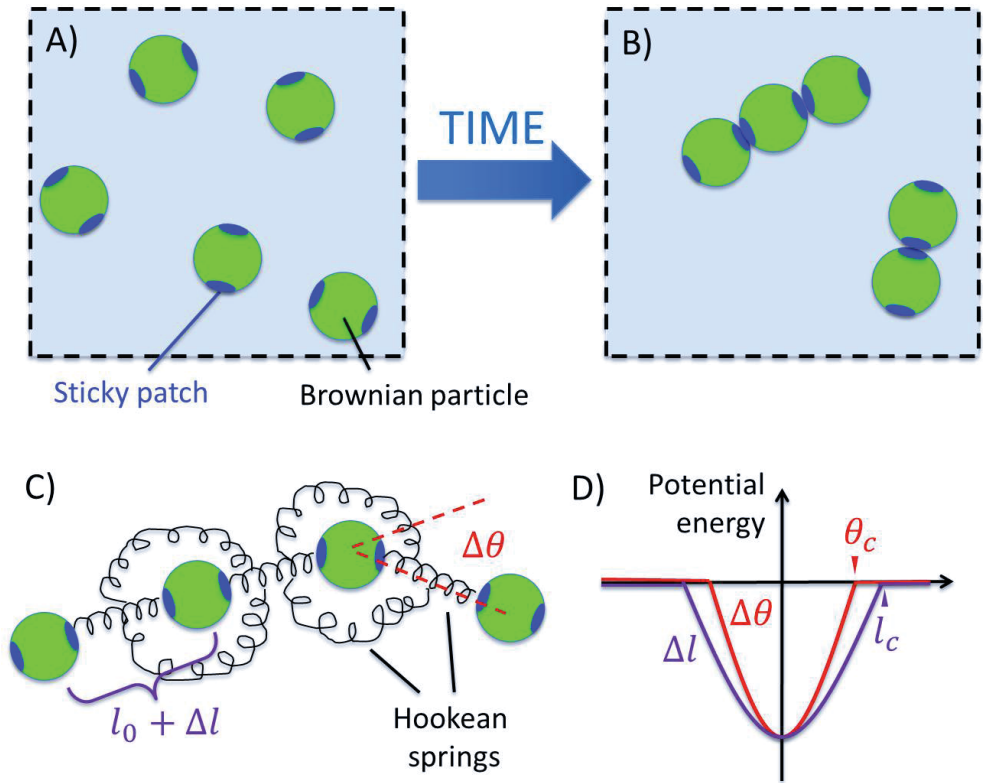}
\end{center}
\caption{(A) A system of spherical particles (beads) with sticky patches on polar ends will self assemble into polymers if the patches are stickiness enough (B). (C) In the minimal model considered here, the interactions within polymer act as Hookean springs between the beads to enforce extensile restriction and rigidity. The size of the beads has been shrunken to show the springs. (D)  The specific potential energy functions governing the deviation in extension $\triangle l$ (solid line) and angle $\triangle \theta$ (broken line) are assumed to be quadratic.}
\label{fig:time}
\end{figure}

\subsection{How to construct a minimal model}
We will start by considering a minimal model of polymerizing monomers as depicted in Fig. \ref{fig:time}A. Namely, the monomers are purely spherical particles (beads) with two sticky patches at two opposite ends. We assume that the beads are in an over-damped environment and so their movement is Brownian. The ``stickiness'' is short-ranged and is quantitatively described by two quadratic energy functions: one controls the distance between the connected beads and the other enforces the rigidity of the resulting polymers (Fig. \ref{fig:time}C and \ref{fig:time}D). We denote the distance between two connected beads by $l_0+\triangle l$. Given any consecutive segment of three monomers in a polymer, if the three  monomers are not co-linear, we denote the angular deviation by $\triangle \theta$ (Fig. \ref{fig:time}C). A polymer  thus consists of a series of beads such that for all consecutive pairs of beads, the absolute values of the deviations, $|\triangle l|$, are smaller than the distance  cutoff $l_c$; and that all deviation angles, $\triangle \theta$, are also less than the cutoff $\theta_c$. The energy function for such a $s$-polymer is then
\begin{eqnarray}
U(\{ \textbf{x} \})=
\sum_{k=1}^{s-1} A\triangle l_k^2+
\sum_{h=1}^{s-2} B\triangle \theta_h^2 -E(s-1),
\end{eqnarray}
where $k$ enumerates the number of distant bonds, $h$ enumerates the number of angular bonds, and $E$ is the binding energy between the patches that promotes aggregation.

Let us now imagine  that at $t=0$, we put these monomers in an inert solvent of  a certain temperature $T$ (Fig. \ref{fig:time}A). The solvent is inert in the sense that their role in the system is purely to provide the thermalizing effects of a heat bath. The whole system is further connected to a much larger thermal bath of temperature $T$ in such a way that heat, but not the beads, can flow back and forth between the system and the large heat bath. In other words, we are investigating the system from the perspective of a canonical ensemble \cite{callen1985thermodynamics}. The setup of this thought experiment corresponds to a typical experimental procedure in which polymerising monomers are first dissolved in an appropriate solvent and then left unperturbed in the course of the experiment. In our case, if the binding energy
is strong enough (i.e., $E$ is large), then we expect that these monomers will self-assemble into polymers (Fig. \ref{fig:time}B). Here, we will assume that the threshold angle $\theta_c$ is small enough that we do not need to worry about the  interactions of distant parts of the same polymer beyond what is already considered in our energy function. In particular, we can ignore the formation of loops in the system.

Although highly simplified, the model presented here is of relevance to some colloidal systems studied experimentally
\cite{wang2012colloids}. But as we will show, the greatest virtue of this minimal model is its analytical tractability.

\subsection{How to deal with the dilute limit}
By  the dilute limit,  we mean that the concentration of solute and the resulting polymers in the system are dilute enough that we can ignore all solute interactions except those that lead to polymerization as described in the previous section. Given this assumption, the free energy density of the overall system can be calculated with the mean-field method. Specifically, we consider a system of $N$ monomers in a volume $V$. The total partition function
can be written in terms of the internal partition function of a single $s$-mer with its first bead's position fixed in space, $z_s$, in the following manner
\cite{lee2009self}:
\begin{eqnarray}
\label{eq:Ztot}
Z_{ tot} =\prod_{s}'\frac{1}{N_s!} \left(\frac{Vz_s}{\Lambda^3}\right)^{N_s},
\end{eqnarray}
where $N_s$ denotes the number of $s$-mers in the system and the prime in the product denotes the number conservation of monomers: $\sum_{s=1}^\infty sN_s =N$. Since we are dealing with ``classical'' (i.e., not quantum mechanical) objects, the kinetic part of the partition function (resulting from momentum integrations) is irrelevant \cite{huang1987statistical} and so $Z_{tot}$ corresponds to the configurational partition function, with $\Lambda$ being an arbitrary constant of dimension length to make $Z_{tot}$ dimensionless.

The denominator $N_s!$ is in (\ref{eq:Ztot}) because the $s$-mers in the system are all indistinguishable and that the free energy is extensive. Note that these polymers are indistinguishable purely because we have chosen not to distinguish them in our analysis, which  is typically the case in experiments \cite{frenkel2014colloidal}.

The total partition function in
 (\ref{eq:Ztot})
follows from a mean-field approximation in the sense that the sequence $\{N_s\}$ is fixed by minimizing the free energy of the system
\begin{eqnarray}
F_{tot} = -k_BT\log Z_{ tot}\ .
\end{eqnarray}
In other words, fluctuations away from the minimising sequence  $\{N_s\}$ are ignored. Such an approximation is expected to be qualitatively correct away from any critical points \cite{callen1985thermodynamics}, which, as we shall see, this system does not possess.

To proceed further analytically, we still need to calculate the $s$-mer partition function $z_s$, which is of the form:
\begin{eqnarray}
\label{eq:zs}
z_s&=&\frac{4\pi l_0^2 {{\rm e}}^{(s-1)\beta E} }{\Lambda^{3(s-1)}}\left(\int_{-l_c}^{l_c} d \triangle l {{\rm e}}^{-\beta A\triangle l^2 } \right)^{s-1}
\left(l_0^2\int_{0}^{\theta_c} d \triangle \theta \sin (\triangle \theta ){{\rm e}}^{-\beta B\triangle \theta^2 } \right)^{s-2}\nonumber\\
\label{eq:zs2}
&=& \frac{4\pi^{3/2} l_0^2{{\rm e}}^{\beta E}}{\Lambda^3\sqrt{\beta A}}  \left(\frac{l_0^2\sqrt{\pi}{{\rm e}}^{\beta E}}{\Lambda^3 \beta^{3/2} \sqrt{A}B} \right)^{s-2},
\end{eqnarray}
for $s>1$, while $z_1=1$. Note that in (\ref{eq:zs}), the factor $4\pi l_0^2$ comes from integrating over the orientation of the polymer given that the first bead is fixed in space, the integrals in the first brackets stem from the longitudinal degrees of freedom and the those in the second brackets from the angular degrees of freedom along the polymer chain. To arrive at (\ref{eq:zs2}), we have taken the limits of integration to infinity, which is legitimate since $\beta A, \beta B$ are typically high where $\beta \equiv k_BT$.

\begin{figure}[h]
\begin{center}
\includegraphics[scale=.7]{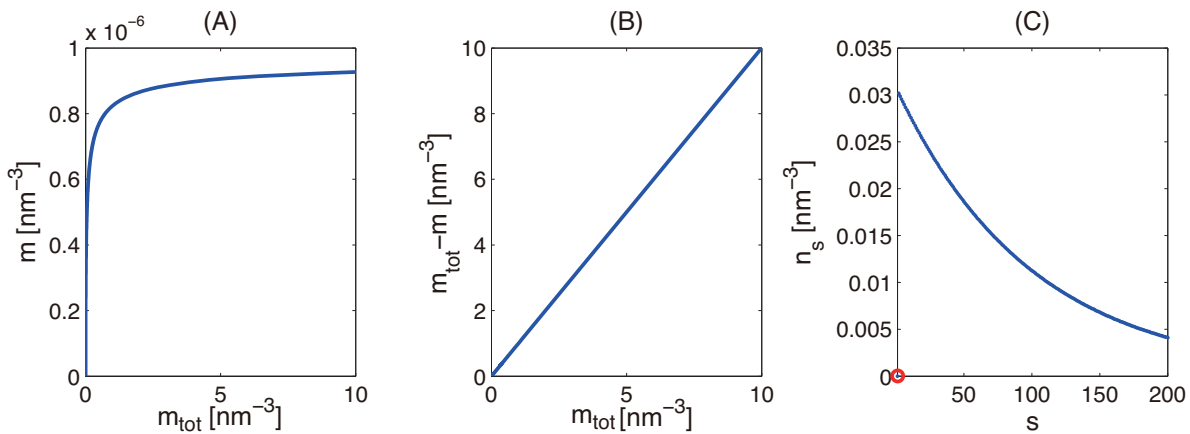}
\end{center}
\caption{Using the parameters in the text, (A) depicts how the monomer concentration changes as monomer concentration increases; and correspondingly for the concentration of the fibrilar species (B). (C) The size (length) distribution of the system at $m_{tot} \simeq 10$nm$^{-3}$. The distribution is exponential but note the discontinuity at $s=1$ (highlighted by the red circle). The unit of the concentrations is nm$^{-3}$.
}
\label{fig:static}
\end{figure}

The total free energy can now be expressed as
\begin{eqnarray}
F_{ tot} &=&-k_BT\sum_s' \left\{ N_s \log \left( \frac{Vz_s}{\Lambda^3} \right) -\log N_s!\right\}
\nonumber\\
&=& -k_BT\sum_s' \left\{ N_s \log \left( \frac{Vz_s}{\Lambda^3} \right) -N_s\log N_s +N_s\right\}
\nonumber\\
\label{eq:f}
&=& \beta^{-1} \left\{N_1 \left[\log N_1 - \log \frac{V}{\Lambda^3} -1 \right]+\sum_s'N_s\left[\log N_s - \log \frac{V}{\Lambda^3} -\chi s-\xi-1 \right]\right\}
\ ,
\end{eqnarray}
 and
\begin{eqnarray}
\xi &=& \log \frac{4 \pi^{1/2} \Lambda^3\beta^{5/2} A^{1/2}B^2}{ l_0^2}-\beta E,
\\
\chi &=& \log \frac{l_0^2\pi^{1/2}}{\Lambda^3 \beta^{3/2} A^{1/2}B}+\beta E
\ .
\end{eqnarray}
 Given  (\ref{eq:f}), we can finally minimise $F_{tot}$ with respect to $N_s$ using the Lagrange multiplier method to enforce the conservation $\sum_s  s N_s =N$. To do so, we minimise the following summation with $\lambda$ being the Lagrange multiplier
\begin{eqnarray}
F_{tot} +\lambda (\sum_s  s N_s -N)
\end{eqnarray}
with respect to the set $\{ N_s\}$, which leads to
\begin{eqnarray}
N_1 &=&\frac{V}{\Lambda^3} {{\rm e}}^\lambda,
\\
N_s &=& \frac{V}{\Lambda^3} {{\rm e}}^{(\chi+\lambda)s+\xi} \quad {\rm for \ } s>1,
\end{eqnarray}
or for $s>1$,
\begin{eqnarray}
\label{eq:Ns}
\frac{N_s}{V} = K\left(\frac{N_1}{V}\frac{1}{m^*_{F}}\right)^s,
\
\end{eqnarray}
where
\begin{eqnarray}
 m^*_{F}&=&\frac{\beta^{3/2} A^{1/2}B}{l_0^2 \pi^{1/2}{{\rm e}}^{\beta E}},
 \\
 K&=&\frac{4 \pi^{1/2}\beta^{5/2} A^{1/2}B^2}{ l_0^2 {{\rm e}}^{\beta E}}.
\end{eqnarray}
 (\ref{eq:Ns}) expresses the $s$-mer concentration $n_s\equiv N_s/V$ in terms of monomer concentration $m=n_1 \equiv N_1/V$. Since $s$ can be as big as we want, by the conservation of mass ($\sum_{s\geq 1} sn_s = N/V\equiv m_{tot}$), we know that $m$ can never exceed $m^*_{F}$ for otherwise the terms in the brackets in  (\ref{eq:Ns}) will blow up with $s$. Indeed, we shall see that $m$ asymptotically approaches $m^*_{F}$ as $m_{tot}$ increases. For this reason, we shall call $m^*_{F}$ the critical fibrilar concentration (CFC) \cite{israelachvili2011intermolecular, lee2009self}. Note that although the system transition from being monomer-dominated to fibril-dominated as $m_{tot}$ increases, it never goes through a phase transition in the thermodynamic sense \cite{huang1987statistical} since the derivatives of the free energy are always continuous. This is also reflected, e.g., by the lack of discontinuities in $m_{tot}$ (Fig. \ref{fig:static}).
In the regime where $m_{tot} \gg m^*_{F}$, $m \simeq m^*_{F}$, and   (\ref{eq:Ns}) shows that the size distribution of polymers is exponential, with the average size given by $\sqrt{m_{tot}/K}$ \cite{lee2009self, cates1990statics, van2006statistical, schmit2011drives}.

Let us now try to substitute in experimentally motivated parameters to see how our model corroborate with observation. Since we are primarily interested in protein aggregation, we take the average size $l_0$ to be 1nm, and the binding energy $E$ to be  25$k_BT$. To estimate the spring constant $A$ and $B$, we make the assumption that each monomer within a polymeric chain has a wriggle room of around 10\% of its size, i.e., $l_c\sim l_0/10$ and $\theta_c \sim 0.1$rad. From this we can estimate $A$ as $100E/l_0^2$ and $B$ as $100E$.  Using these parameters, we find that $m^*_{F} \simeq 9.8 \times 10^{-7}$nm$^{-3}$ or around 9.8$\mu$M. The corresponding fibrillation behaviour of this system is shown in Fig. \ref{fig:static}.

With regard to experimental observation, the predicted exponential length distribution seems to deviate from some experimental studies \cite{rogers2005measuring, xue2009amyloid}. Van Raaij {\it et al.} has interpreted the observed peaks as a
result of the finite resolution of the atomic force microscopy
imaging and length measurement procedure \cite{van2008concentration}. Besides this
explanation, it is also known that it can take on the order
of months for mature fibrils to form \cite{morel2010environmental}. Therefore, the
appearance of the peaks observed may also reflect the fact
that the self-assembled systems have not yet reached thermal
equilibrium.

\subsection{How to generalize to the semi-dilute limit}
In the dilute limit, mutual interactions  between the solutes beyond the polymerizing interactions are ignored. What if we now increase the concentration of the solute so that such an approximation is no longer valid. This takes us to the semi-dilute limit. We will first discuss the simplest and ubiquitous type of interactions: volume exclusion interactions.

\subsubsection{Pure volume exclusion interactions}
To take steric interactions into account, we start again with the non-interacting free energy density (see (\ref{eq:f})):
\begin{eqnarray}
f_0 = \beta^{-1} \int d s n(s)\left[\log n(s) + \log {\Lambda^3} -\chi s-\xi-1 \right],
\end{eqnarray}
where we have ignored the monomeric contribution, and pass to the continuum description in $s$ since we are primarily interested in the fibril-dominant regime ($m_{tot} \gg $CFC). We then add to $f_0$ the following interaction term:
\begin{eqnarray}
f_{\rm int} = \int d s d s' n(s) n(s') B(s,s')
\ .
\end{eqnarray}
In the above equation, $B(s,s')$ is the second virial coefficients corresponding to the steric interactions of two semi-flexible polymers of length $s$ and $s'$, and is of the form \cite{van1994growth}:
\begin{eqnarray}
\label{eq:B}
B(s,s')=\frac{2\pi}{3}l_0^3+\frac{\pi(s+s')l_0^3}{2}+2ss'l_0^3|\sin \phi|,
\end{eqnarray}
where $\phi$ is the angle between the two polymers. To deal with the additional variable $\phi$, we ask ourselves what would happen to the system given the steric interactions. From the physics of liquid crystals \cite{prost1995physics, edwards1986theory}, we expect that as the polymer concentration increases, the system can become nematic, i.e., the semi-flexible polymers will be aligned.

\begin{figure}[h]
\begin{center}
\includegraphics[scale=.6]{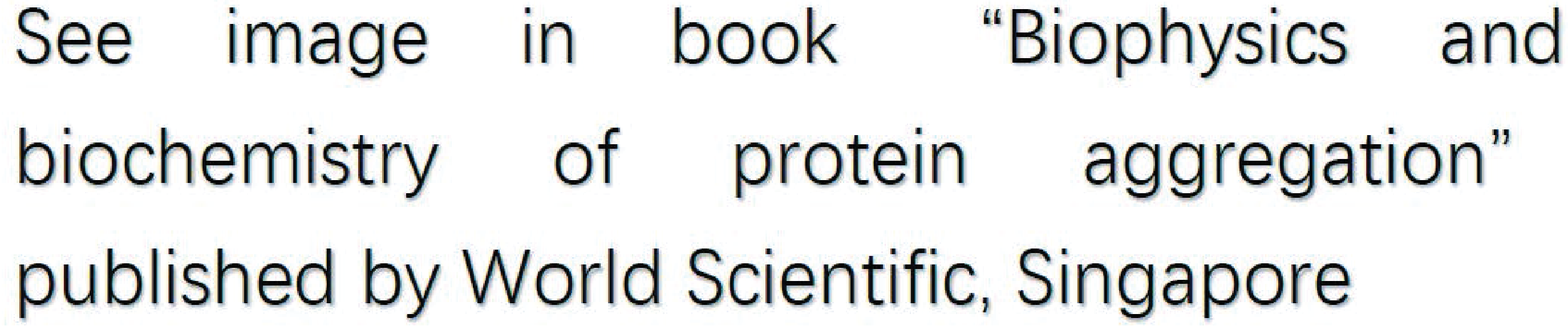}
\end{center}
\caption{Glass vials (1 cm wide) with hen lysozyme fibril containing solutions imaged between crossed polars. Concentrations are indicated in mM. The lit up regions correspond to the nematic phase. Image is taken from \cite{corrigan2006formation} with copyright permission.
}
\label{fig:iso}
\end{figure}

In other words, we anticipate that similar to liquid crystals \cite{edwards1986theory}, as the solute concentration increases, the system will
 first go through a phase separation where the regions of the system can be partitioned into two phases, one nematic and the other isotropic. If the concentration increases further, the system will become fully nematic. This is indeed what is observed experimentally in, for instance, a system of fibrilising hens lysozyme \cite{corrigan2006formation} (Fig. \ref{fig:iso}). We will now incorporate this expected picture into our free energy minimisation. Specifically, we will consider both free energy densities in the isotropic and nematic phases, $f_{ I}$ and $f_{ N}$ respectively. For $f_{ I}$, we can simply add to  (\ref{eq:f}) the average over the angle that two randomly oriented semi-flexible make (since the system is isotropic) in  (\ref{eq:B}), hence
 \begin{eqnarray}
 f_I = f_0 (\{ n_I(s)\} ) +
\int ds ds' n(s) n(s') \left[\frac{2\pi}{3}l_0^3+\frac{\pi(s+s')l_0^3}{2}+\pi ss'l_0^3 \right] \ .
 \end{eqnarray}
 Using again the Lagrange multiplier method to enforce protein number conservation, one finds that the distribution $n_{ I}(s)$ that minimises $f_{ I}$ is exponential as before, although the mean size is now $\sqrt{m_{I}{{\rm e}}^{8 \psi_I/3} /K}$, where  $m_{I}$ is the total protein concentration and
  $\psi_I$ is the volume fraction of proteins in the isotropic phase \cite{van1994isotropic}.

 In the nematic phase, the picture is more complicated. Since the polymers can elongate, it was found that the flexibility of the polymer has to be taken into account in order to stop the unrealistic lengthening of the polymers in the nematic phase \cite{odijk1987effect}.
 Here, we will again quote the results in \cite{van1994isotropic}, in which it was found that by minimising the free energy in the nematic phase, the length distribution is again exponential, and the mean size is now approximately $\sqrt{m_{N}{{\rm e}}^{4\psi_I} /K}$. Here, $m_{N}$ is the total protein concentration in the nematic phase. These results are in the regime where the mean polymer length is much greater than the persistence length of the polymer.
%

 Minimizing $f_I$ and $f_N$ separately is not the whole story since there is also the possibility of phase separation. Namely, the volume of the system can be partitioned into isotropic and nematic regions. In the thermodynamic limit, we usually ignore the surface energy coming from the interface separating the isotropic and nematic regions. The minimisation is thus performed on the following free energy density:
 \begin{eqnarray}
 \label{eq:ftot}
 f_{tot}(v_{tot},m_{tot}) =\frac{v_If_I(m_I)+v_Nf_N(m_N)}{v_{tot}},
 \end{eqnarray}
 where $v_I (v_N)$ is the total volume of the isotropic (nematic) regions and $c_I (c_N)$ is  solute concentration in the isotropic (nematic) regions. The conservation laws are therefore $v_I+v_N=v_{tot}$ and $v_Im_I+v_Nm_N=v_{tot}m_{tot}$. Minimising $f_{tot}$ with respect to $v$ and $m$ finally enables us to confirm the expectation we had from the beginning. Namely, as solute concentration increases, regions of nematic phase appear in the system and co-exist with the isotropic phase. The polymer length distributions in both phase remain exponential, although the average length in the nematic is higher than that in the isotropic region \cite{lee2009elongation}. As the solute concentration increases further, regions of isotropic phase disappear and the whole system will be in the nematic phase. As mentioned, these theoretical findings are corroborated by experimental study on  self-assembling hens lysozyme \cite{corrigan2006formation, lee_isotropic09} (see Fig. \ref{fig:iso}).

\subsubsection{With lateral interactions}
We have seen how the isotropic-nematic phase transition in a system of semi-flexible polymers is typically accompanied by phase separation. In this section, we will show that if the polymers are mutually attractive (beyond the end-to-end binding that leads to the joining of the two polymers), then the tendency for phase separation is even stronger. Specifically, using again our minimal model system shown  in Fig. \ref{fig:time}, we imagine that besides the strong directional attractive interactions schematically depicted by the red (dark grey) patches, the beads are weakly attractive($\sim k_BT)$ towards each other, i.e., the green (light grey) areas of the beads are also weakly sticky.

To appreciate conceptually how interacting polymers behave at thermal equilibrium, the Flory-Huggins theory is a good starting point since it is conceptually simple and can be easily analysed numerically. Here, we will ignore the rigidity of the polymers and  focus solely on the effects of attractive interactions on the system's phase behaviour. Using the lattice model where each lattice site can be either occupied by one solvent molecule and by one monomer, the following Flory-Huggins free energy density of {\it mixing} can be derived \cite{barrat2003basic}:
\begin{eqnarray}
\label{eq:fh}
f_m(\phi) = \frac{1}{\bar{n}}\phi \ln \phi +(1-\phi) \ln (1-\phi) + \chi \phi(1-\phi),
\end{eqnarray}
where $\phi$ is the volume fraction of the polymers in the system and $\bar{n}$ is the number of monomers in the polymer. We note that in the above formula, all polymers are of the same length, i.e., we have ignored the disperse length distribution here for simplicity.  The parameter $\chi$ summarises the interactions of the monomers and is of the form
\begin{eqnarray}
\chi\equiv \frac{z}{2k_BT} [2e_{ms}-e_{mm}-e_{ss}],
\end{eqnarray}
where $e_{mm},e_{ss},e_{ms}$ are the interaction energy for the monomer-monomer, solvent-solvent, and monomer-solvent pairs on the lattice, respectively.
Here, we assume that the monomers are weakly attractive ($e_{mm} \sim -k_BT$) and for simplicity, we  set $e_{ss}=0=e_{ms}$. Moreover $z$ is the coordination number of the lattice, which for a 3D cubic lattice is 6. As a result, the interaction parameter $\chi$ is $-3e_{mm}/k_BT$.

\begin{figure}[h]
\begin{center}
\includegraphics[scale=.6]{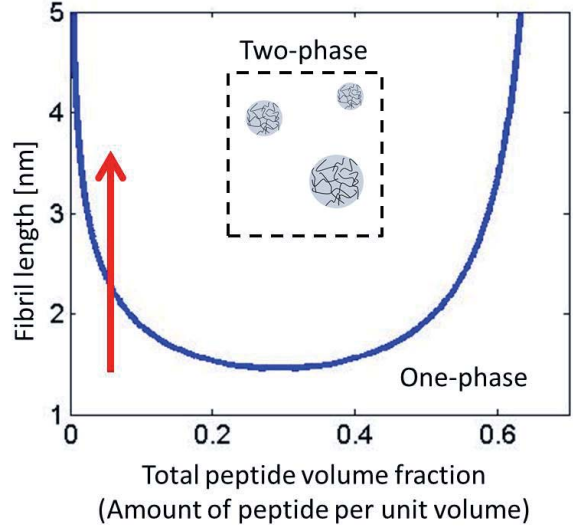}
\end{center}
\caption{A typical phase diagram of a phase separating polymeric system according to the classic Flory-Huggins theory \cite{doi1996introduction}. Phase separated polymer drops (inset) form in the two-phase region is bounded by the blue (dark grey) curve. At fixed peptide volume fraction, the tendency for the system to phase separate increases as the self-assembled fibrils elongate (indicated by the red (light grey) arrow).
}
\label{fig:ps}
\end{figure}

The first two terms in  (\ref{eq:fh}) promotes phase separation, while for positive $\chi$, $f_m$ becomes concave down when $\chi$ is large enough, this is a signature of phase separation, which means that we will need to minimise the total free energy by considering the possibility of having the system partition into two parts of distinct phases as in the previous section (see  (\ref{eq:ftot})). A typical phase diagram is shown in Fig. \ref{fig:ps}.

\begin{figure}[h]
\begin{center}
\includegraphics[scale=.6]{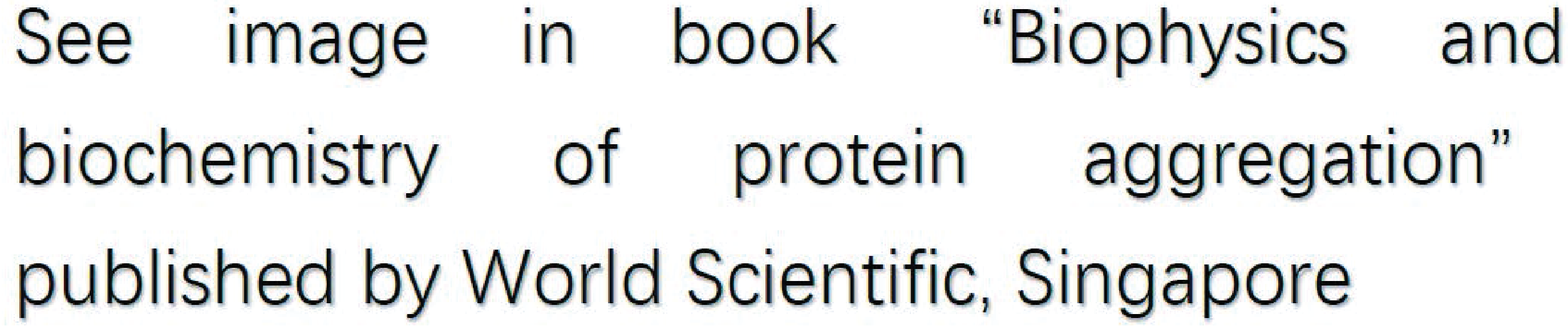}
\end{center}
\caption{
(A) Section through
a tomogram of a yeast cell in which amyloid fibrils formed from Sup35 are close to the cell membrane.  The nucleus is outlined in magenta
and the vacuole in brown. (B) Rendered 3D model from
serial tomograms of this cell. Amyloid fibrils are in green;
membrane, blue; nucleus, magenta; vacuole, brown; mitochondria,
purple; and large complexes (presumably ribosomes)
as gray dots.  (C) Section through of a cell in which the aggregate of amyloid fibrils is of a form of a drop (green). (D) Rendered 3D model of
the dot from serial tomograms. (Coloring as in B.) This figure is taken from \cite{saibil2012heritable} with copyright permission.
}
\label{fig:yeast}
\end{figure}

The key feature here is that as the polymers elongate, the tendency to phase separate gets stronger even if the polymers are only weakly attractive towards each other. Experimentally, it is observed that many amyloid fibrils aggregate in solution and thus {\it in-vitro} and {\it in-vivo} phase separation of amyloid fibrils may be expected to occur. Indeed, the observed clustering of sup35 fibrils in the cytoplasm of yeast cells may be a signature of such phase separation (Fig. \ref{fig:yeast}).

\subsection{How to incorporate oligomers}
We have so far focused on the behaviour of fibrilising system in which there are only monomers or fibrils. In the case of amyloid fibrils, the situation is more complex. Indeed, mounting evidence has indicated that proteins in the monomeric form and oligomeric form (potentially amorphous aggregates of tens of proteins), instead of proteins in the fibrillar form, are predominantly responsible for cell death  \cite{lambert1998diffusible}. In this section, we will incorporate the oligomeric species into our analysis. To model the presence of oligomers, we borrow the treatment of spherical micelles formation in a solution of surfactants \cite{israelachvili2011intermolecular}. Specifically, we assume that the monomers can aggregate together to form an amorphous cluster. However, there is an optimal number of monomers, $W$, in the cluster in the sense that the corresponding cluster partition function is greatest. In this system, a monomer can be classified into three categories: (1) monomeric, (2) part of an oligomer (which we call a micelle), and (3) part of an fibril. The total partition function in this system can be written as \cite{lee2009self}:
\begin{eqnarray}
\label{eq:Ztot2}
Z_{ tot} =\prod_{s,p}'\frac{V^{N_1}}{\Lambda^{3N_1} N_1!} \frac{(Vz^{(f)}_s)^{N_s}}{\Lambda^{3N_s}N_s!}\frac{(Vz^{(m)}_{p})^{M_{p}}}{\Lambda^{3M_p} M_{p}!},
\end{eqnarray}
where $z^{(f)}_s$ and $z^{(m)}_p$ are the internal partition functions for the fibrillar and micellar species. We have also singled out the monomeric contribution to the partition to highlight the fact that there are three different species in the system. We now model the micellar partition function as
\begin{eqnarray}
z^{(m)}_p = \delta_{pW}\left(\frac{\alpha}{\Lambda^3}\right)^{W-1}{{\rm e}}^{WE_m}
\ ,
\end{eqnarray}
where $\alpha$ is of the order of the dimension of the monomer. Specifically, all micelles are assumed to be of size $W$ for simplicity  and the clustering is driven by the binding energy $E_m$ per monomer.  If we ignore the fibrillar species for the time being, then using again the Lagrange multiplier method as before, we find that we can again relate the micellar concentration to the monomer concentration $m$:
\begin{eqnarray}
m_W = \frac{1}{\alpha}\left( \frac{m}{m^*_M}\right)^W
\ ,
\end{eqnarray}
where $m^*_M = (\alpha {{\rm e}}^{\beta E_m})^{-1}$. For $W \sim {\cal O}(10)$, we can  see that $m$ is bounded above at around $m^*_M$, i.e., most monomers are in the micellar form at $m_{tot} >m^*_M$. On the other hand, for $m \ll m^*_M$, $m_K$ is negligible.
 For this reason, we call $m^*_M$ the critical micellar concentration (CMC).

\begin{figure}[h]
\begin{center}
\includegraphics[scale=.6]{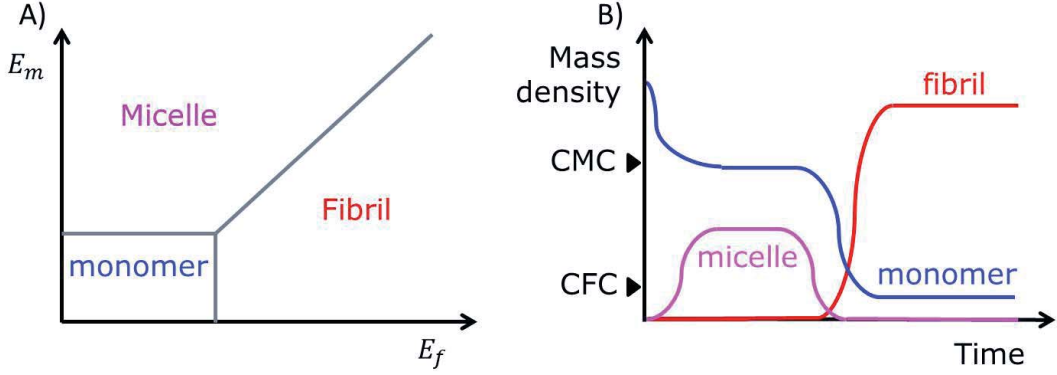}
\end{center}
\caption{(A) At fixed $m_{tot}$, the system can be dominated by distinct species depending on the strength of the fibrilar binding energy $E_F$ and the micellar binding energy $E_M$. (B) A schematic of a potential nucleation pathway of fibrillation proposed in \cite{lomakin1997kinetic}. Starting with a pool of monomers, micelles form quickly due to  the fast (potentially diffusion-limited) formation kinetics. within the micelles, the protein concentration is high and thus facilitates the slow nucleation of nuclei of fibrils. As fibrils elongate, the monomers in the system will be eventually depleted below the CMC and thus the micelles will disappear in the system.
}
\label{fig:oligomers}
\end{figure}

If we now incorporate the fibrilar species into the picture, the system is effectively partitioned into multiple regimes: if $m_{tot}$ is smaller than both the CMC and the CFC, then the system is dominated by $m_{tot}$. If the CMC is lower than the CFC and $m_{tot}>$ CMC, then the micellar species will dominate the system; while if the CFC is lower than the CMC, fibrils will be dominant if $m_{tot} >$ CFC \cite{lee2009self} (Fig. \ref{fig:oligomers}A). Note that even though if the CFC is lower than the CMC, the micelles would still be transiently present in the system, and their existence may have important implications in the fibrillation kinetics as proposed in \cite{lomakin1997kinetic} (Fig. \ref{fig:oligomers}B). A similar model has also been recently studied using molecular dynamics simulation in \cite{Saric2014}.

\subsection{How to incorporate internal monomeric structures}
Another natural generalisation of our minimal bead-and-stick model is the incorporation of the internal monomeric structure into our analysis. A peptide can in general be in multiple states, e.g., in the form of a beta sheet, random coil, alpha helix. In the specific case of A$\beta$ peptides, the monomeric state is in the random-coil configuration, while the peptides are predominately in the beta-sheet state in the fibrillar form \cite{petkova2002structural}. In terms of our minimal model, we can account for this modification by assigning a nonzero value to the monomeric partition function $z_1^{(R)}= \gamma >0$ corresponding to the random coil state in (\ref{eq:Ztot}); while for monomers in the beta-sheet form, the monomeric partition function $z_1^{(\beta)}$ is again set to one. Here, $\gamma$ is positive means that the monomer in the random coil state is preferred over the beta sheet state in the fibrillar form, which originates from the fact that the random coil is entropically more favourable.
At thermal equilibrium, we expected have $m^{(\beta)} = {\rm e}^{-\gamma} m^{(R)}$. The total monomer concentration $m$ is therefore $m^{(\beta)} + m^{(R)}=m^{(R)}(1+\rm{e}^{-\gamma})$. Incorporating the fibrillar species into the picture and writing the concentration of $s$-mers in terms of that of the monomer as done in (\ref{eq:Ns}), we have
\begin{eqnarray}
n_s = K\left(\frac{m^{(R)} }{{\rm e}^{\gamma}{m}^*_F}\right)^s=K\left(\frac{m}{\hat{m}^*_F}\right)^s
\end{eqnarray}
where $\hat{m}^*_F$ is $(1+{\rm e}^{\gamma}) m^*_F$ (see (\ref{eq:Ns})). In other words, most of qualitative analyses as before remains the same, except now the critical fibrillar concentration is increased by the factor $(1+\rm{e}^{\gamma})$, which signifies that the monomeric concentration $m$ is generally increased when the random coil-beta sheet transition \cite{hong2008statistical1, hong2008statistical2} is taken into account at the monomeric level. We note that a more detailed analysis of the effects of the degree of freedom from the monomeric conformation can be found in \cite{schreck2013statistical}.

\subsection{How to incorporate multiple fibril morphologies}
\begin{figure}[h]
\begin{center}
\includegraphics[scale=.5]{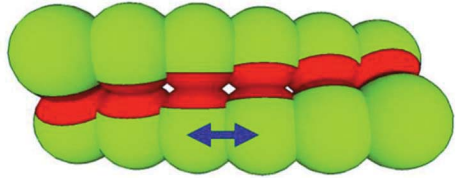}
\end{center}
\caption{A schematic of a twisting fibre consisting of two filaments. In this model, besides the axial bonds (blue arrows) along the longitudinal direction of the fibre, there are lateral bonds (red patched) that bind the two filaments together.
}
\label{fig:fiber}
\end{figure}

Besides the possibility of micelle formation, real biopolymers, and amyloid fibrils in particular, are likely to consist of two or more filaments. In addition, multiple morphologies may co-exist in the system \cite{usov2013polymorphism}. The simplest way to model such a system is to imagine that each bead in  Fig. \ref{fig:time}A also possesses a sticky patch on the equator. If the new ``patch on the side'' is small in area, then a two-filament fibril will form naturally (Fig. \ref{fig:fiber}). In fact, many amyloid fibrils seemed to consist of filaments twisted together, which could be incorporated into our minimal model by twisting the location of the side patch along the axial direction (Fig. \ref{fig:fiber}).
The formalism employed so far can again be generalised to consider this system. Interestingly, as far as  minimising the total free energy is concerned, the only effects of the lateral association of filaments are to double the capping energy $E$, where the factor $2$ comes from the additional axial bond due to the bundling of the two filaments \cite{lee2009elongation}. Because of this increase of capping energy {\it via} bundling, the species that dominate the system will always be the two-filament polymers. So how does this theoretical predictions square with experimental observations that multiple morphology exist? The simplest resolution again points to the conclusion that under typical experimental conditions, the self-assembly of amyloid fibril has not yet reached the thermal equilibrium state.

\subsection{Summary}
Here we will summarise the key conclusions one can draw from statistical mechanics with regards to biopolymer self-assembly.
\begin{enumerate}
\item
The existence of a critical fibrillar concentration (CFC) below which fibrillar mass is negligible. The emergence of such a critical concentration comes purely from the fact that the aggregates concerned consist of a large number of monomers.
\item
Above the CFC, the length distribution of the fibrils is exponential. This remains so even in the semi-dilute limit if the fibril-fibril interactions are purely steric.
\item
When micellar or other oligomeric species are taken into account, each distinct type of aggregates will have their specific critical concentrations, and the one with the lowest critical concentration will have the dominating mass density at high protein concentration (Fig. \ref{fig:oligomers}).
\item
If fibrils of distinct widths exist, the mass density of the widest fibrils will dominate the system at thermal equilibrium.
\item
If the fibrils exhibit attractive lateral interactions, there will be a strong tendency for phase separation of fibrils to occur.
\end{enumerate}

We note that the existence of CFC is a well established experimental observation. However, experimentally measured fibrillar length distributions seem generally to deviate from the predicted exponential distribution. In addition, multiple morphologies of fibrils seem to co-exist in  typical experimental conditions. These observations indicate that protein amyloid formation does not reach thermal equilibrium at the typical experimental timescale. Therefore, to account quantitatively for the experimental observation, we generally need a kinetic description of amyloid fibrillisation tailored for the specific experimental condition.

\section{Kinetics of amyloid fibrillation}
In the first part, we have presented a general review on the thermodynamics of amyloid fibrillation. However, due to the lack of a mature non-equilibrium thermodynamics theory, currently we can only deal with the equilibrium states and the phase transition between them. In order to understand many unsolved puzzles concerning with the time evolution of an amyloid system, such as how fibrils grow, how they replicate, how they interact with cells and be cytotoxic, we need to turn to a kinetic theory. The mathematical modeling based on chemical mass-action equations provides us a unified framework as well as a quantitative linkage between experimental observations and underlying molecular mechanisms. Fruitful results and interesting physical insights have been obtained based on this formulation in the past several years. In what follows, we will focus on two aspects: one is a systematic exploration of the kinetic formulation of amyloid fibrillation, including both molecular mechanism basis and model analysis (Fig. 8); the other is the mathematical foundation of kinetic modeling as well as its linkage with experimental facts.


\begin{figure}[ht]
	\centering
		\includegraphics[scale=0.29]{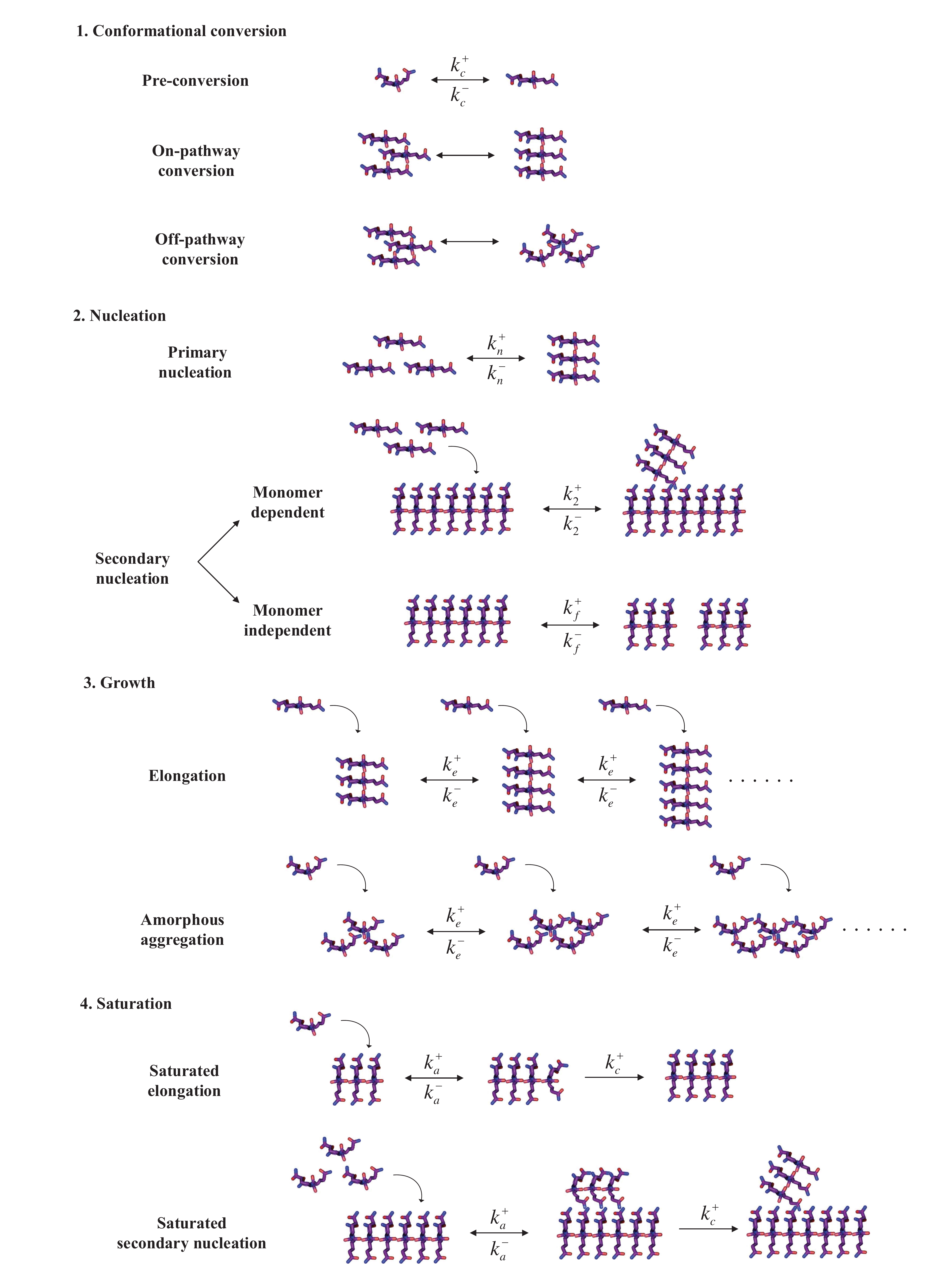}
		\caption{A cartoon depicting various microscopic mechanisms of amyloid fibrillation discussed in the current paper.}
		\label{Fig:mechanism}
\end{figure}

\subsection{How fibrils grow}
When examining a given amyloid system, a first question coming to mind usually would be how amyloid fibrils are able to grow? Regardless of various possible explanations for other self-assembling phenomena in nature \cite{whitesides1990wet,ulman1996formation,schreiber2000structure,ulman2013introduction}, we are really astonished at the fact that there is one truly simple answer valid for almost all amyloid fibrils. That is elongation, the process of which incorporates free protein molecules (or monomers as stated in literature) into existing fibrillar aggregates by monomer association and dissociation at fibril ends in a linear sequential way. Elongation not only is a geometrical consequence of the intrinsic one-dimensional structure of amyloid fiber, but also has a deep root in both thermodynamics and kinetics, that is monomer association is more favorable than that of oligomers in general \cite{oosawa1975thermodynamics, collins2004mechanism}. We will come back to this point later.

As an elementary step in amyloid fibrillation, elongation has been paid great attention to in the past studies. Pioneer works dated back to Oosawa and his colleagues for their preliminary examination on actin formation in the late 1950s \cite{oosawa1959g}. They are among the first ones who wrote down the explicit reaction schemes and rate laws for protein self-assembling, \textit{i.e.} once formed by nucleation processes (which will be discussed in next section), actins will extend or shrink by association or dissociation of monomeric units at one end mostly. They further verified that the initial rate of actin growth varies linearly with the monomer concentration, which implies that monomeric, rather than oligomeric, subunits have been added to actin filaments and make them to grow \cite{oosawa1975thermodynamics}, since the latter will give rise to a nonlinear dependence on the monomer concentration.

The linear dependence of elongation rate on the monomer concentration has been further investigated by Collins \textit{et al.} through a combination of kinetic modeling and single-molecule fluorescence measurements on the NM domain of yeast prion protein Sup35 \cite{collins2004mechanism}. Again, the initial rate of fiber growth was found to be directly proportional to the concentration of soluble NM over a range of $0\sim 1 \mu M$. However, at a higher NM concentration ($>10\mu M$), the rate of fiber growth shows a weaker-than-linear law. The reason for this is believed to be ``a conformational rearrangement of NM after binding to fiber ends becomes rate limiting at high NM concentration'' \cite{collins2004mechanism}, a specific case of the saturation phenomenon. A quantitative treatment will be presented in the section of saturation.

Recently, Knowles and his colleagues developed a novel technique to measure the rate of fiber elongation directly \cite{knowles2007kinetics, buell2010frequency}, which to some extent avoids the indistinguishability of several different fibrillation processes presented simultaneously in traditional methods. According to their setup, prepared fragments of amyloid fibrils have been attached to the surface of a quartz transducer. So that, once new monomers add to the fibrils, the resulting nanogram mass changes will be monitored through a shift in the resonance frequency of the quartz oscillator. Again, two different regimes in the fiber elongation rate are highlighted: an initial linear dependence on the monomer concentration, and a subsequent saturation of the growth rate at high monomer concentrations.

In literature, the possibility of fiber elongation through oligomer addition has been discussed from time to time \cite{serio2000nucleated,souillac2003structural,caughey2003protofibrils,bitan2003amyloid}. For example, Serio \textit{et al.} proposed a oligomer-based mechanism called ``Nucleated Conformational Conversion'' \cite{serio2000nucleated}, in which ``structurally fluid oligomeric complexes appear to be crucial intermediates in de novo amyloid nucleus formation'' and ``rapid assembly ensues when these complexes conformationally convert upon association with nuclei'', to explain the assembling of prion protein Sup35. However, since in general ``monomer addition is more rapid and efficient'' than oligomer addition \cite{collins2004mechanism}, nowadays a common conclusion has been reached that fiber elongation has a first-order concentration dependence on both monomeric and fibril species, revealing a bimolecular mechanism of growth through monomer addition to both ends of existing fibrils.

Now we are going to formulate above discussions into a quantitative model. If only the elongation process is included, it is straightforward to show that the mass concentration and number concentration of aggregates (see their mathematical definitions and physical meanings in section of how to quantify fibrillation kinetics) evolve according to following equations:
\begin{eqnarray}
&&\frac{d}{dt}P=0,\label{elongationP-1}\\
&&\frac{d}{dt}M=k_e^+mP-k_e^-P,\label{elongationM-1}
\end{eqnarray}
where $k_e^+$ and $k_e^-$ denote the reaction rate constants for monomer association and dissociation respectively. Clearly, the first term on the right-hand side of (\ref{elongationM-1}) represents the desired process of fiber elongation by monomer addition at fibril ends (here the geometrical factor 2 is not directly written out); while the second one is the corresponding inverse process, whose necessity lies on ``the maintenance of monomer pool'' in the equilibrium state \cite{nuallain2005monomer}.

It is noted that, according to (\ref{elongationP-1}), the number concentration is conserved, \textit{i.e.} $P(t)=P(0)$. This is a significant feature of elongation, in contrast to all kinds of nucleation processes which will be introduced in the following section. Towards the mass concentration, obviously its growth rate is maximal at the beginning of the reaction and then decreases monotonically. Actually, we have
\begin{eqnarray}
M(t)=(M_0 - \epsilon)e^{-\kappa t}+\epsilon,
\end{eqnarray}
where $\kappa=k_e^+P_0$, $\epsilon=m_{tot} - k_e^-/k_e^+$, $M(0)=M_0$ and $P(0)=P_0$ are the initial mass and number concentrations of aggregates. Accordingly, the half-time and apparent fiber growth rate (see section on how to quantify fibrillation kinetics for their definitions) are given by
\begin{eqnarray}
&&t_{1/2}=\frac{\ln2}{\kappa}\propto P_0^{-1},\\
&&k_{app}=\frac{\epsilon+M_0}{2m_{tot}}\kappa\propto P_0.
\end{eqnarray}
Here, both $t_{1/2}$ and $k_{app}$ are only concerned with the number concentration of seeds $P_0$ rather than the mass concentration $M_0$. This is another significant feature of the elongation process. Contrarily, as we will show, the speed of secondary nucleation generally depends on the mass concentration of seeds $M_0$; while that of primary nucleation does not rely on seeds at all. Finally, from the static solution $M(\infty)=\epsilon$, we can exactly see that the inverse reaction -- monomer dissociation is essential to maintain a genuine thermodynamic equilibrium as well as an observable ``monomer pool'' at the end of reaction \cite{nuallain2005monomer}.

In (\ref{elongationM-1}), the rate constants for monomer association and dissociation are assumed to be independent of fibril length. An indirect support comes from the argument that they are ``characterized by the local interaction of a monomer with the end of a fibril'', so that ``the total length of the fibril is likely to introduce only a minor effect through the reduction of the diffusive encounter rate when the size of the fibril increases'' \cite{buell2010frequency}. Under the assumption of fibril length independence, there are plenty of studies attempting to determine exact values of the rate constants for monomer association and dissociation. For example, by AFM analysis Collins \textit{et al.} reported the rate constant for fiber elongation of yeast prion protein Sup35 is approximately $k_e^+\approx2\times 10^{5} M^{-1}s^{-1}$ at a soluble NM concentration of $2.5\mu M$ and can be two times bigger at higher concentrations \cite{collins2004mechanism}. Knowles \textit{et al.} found an elongation rate of $(9.2\pm 0.3)\times10^3 M^{-1}s^{-1}$ for insulin at a concentration of $0.17mM$ by quartz crystal microbalance, equivalent to one molecule attaching every $3.1\pm1.2s$ on average \cite{knowles2007kinetics, buell2012rate}.  At $2.5mg/ml$ $A\beta_{25-35}$ peptide, Kellermayer \textit{et al.} determined $k_e^+\approx10^6 M^{-1}s^{-1}$ and $k_e^-\approx10s^{-1}$  through scanning-force kymograph, given the equilibrium monomer concentration to be $10\mu M$  \cite{kellermayer2008stepwise}. In a similar way, referring to a critical concentration of $\alpha$-synuclein at $2\mu M$, Pinotsi \textit{et al.} \cite{pinotsi2013direct} gave an average growth rate of $k_e^+\approx(1\pm0.375)\times10^3 M^{-1}s^{-1}$ as well as a lower bound for the average dissociation rate $k_e^-\approx(2.0\pm0.8)\times10^{-3} s^{-1}$ by two-color single-molecule localization microscopy.

From above far-from-complete list, we can clearly see that the rate constant for fiber elongation varies for at least three orders of magnitude from protein to protein and could also be influenced by the pH condition, salt concentration, temperature and \textit{etc.} \cite{pinotsi2013direct}. This fact to some extent reveals the intrinsic high complexity and heterogeneity of fiber elongation processes. Actually, for the growth of individual filament, an intermittent, stop-and-go behavior has been widely confirmed \cite{ferkinghoff2010stop, wordehoff2015single}, which shows fiber elongation operates ``in a way analogous to the landscape models of protein folding defined by stochastic dynamics on a characteristic energy surface'' \cite{lee2009elongation, buell2010frequency}. Therefore, the current single-reaction-rate-based picture turns out to be a rough average of various underlying stochastic processes concerning with the fiber growth. Readers should be aware of this point.


\subsection{How fibrils replicate}
Generally speaking, the fibrillation of amyloid proteins is constituted by two aspects: nucleation and growth. In the previous section, we have shown how elongation provides a simple way to let fibrils grow. But we still lack a knowledge on where those templates (or seeds) for elongation come from and how they evolve during the fibrillation procedure. As far as we know, there are at least three basic processes contributing to the generation of new seeds, \textit{i.e.} primary nucleation, surface-catalysed secondary nucleation and fragmentation. There is a long story for the study of primary nucleation in crystal and liquid formation. Surface-catalysed secondary nucleation and fragmentation present two alternative ways to bypass the slow seeding procedure of primary nucleation by making use of existing fibrils in the system, so they are also named as ``secondary nucleation'' in literature. A major difference between them is that the former dependents on the monomer concentration, while the latter does not.

Secondary nucleation plays a key role in the fibrillation of amyloid proteins \cite{knowles2009analytical,gillam2013modelling}. In the presence of secondary nucleation, the fibrillation speed will be dramatically accelerated and time courses for fiber mass concentration will be changed into a sigmoidal shape with prominent lag phase. Furthermore, in contrast to primary nucleation and surface-catalysed secondary nucleation, fragmentation will not only cause a global change of the fiber length distribution, but also largely enhance the cytotoxicity by generating more harmful oligomers. Detailed discussions could be found in following sections correspondingly.

\subsubsection{Primary nucleation: rate-limiting step}
For different amyloid proteins under different conditions, time courses for fibrillation may vary considerably from each other, but in general a sigmoidal-like behavior with a prominent lag phase is observed if initial proteins are all in monomeric form \cite{knowles2011observation}. Furthermore, by introducing certain amount of pre-seeded filaments into the system at the very beginning of the reaction, the lag phase can be completely removed \cite{cohen2012macroscopic}. This evidence reveals that amyloid fibrillation involves a process called primary nucleation, in which nuclei of aggregates are formed from monomeric proteins directly. In contrast, the formation of amorphous aggregates, which often acts as competitive pathways against regular amyloid fibrils, is generally believed to follow a non-nucleated polymerization (or random polymerization) \cite{kodaka2004interpretation, kodali2007polymorphism, powers2008mechanisms}.

The idea of primary nucleation has been extensively explored in various natural phenomena, such as the formation of snow flakes, clouds and bubbles, the crystallization of mineral, metal, protein and DNA, \textit{etc.} \cite{van1889continuity,becker1935kinetic,frenkel1955kinetic,lothe1968concentration,reiss1968translation,kikuchi1969translation,shi2012van}. It is the first step in the formation of either a new thermodynamic phase or a new structure via self-assembly, and typically determines how long we have to wait before the new phase or self-organised structure appears. According to whether the process is catalyzed by particles of foreign substance (like surface and substrate) or not, primary nucleation could be further divided into two categories: homogeneous nucleation and heterogeneous nucleation. The latter usually occurs much more often and faster than the former. This behavior could be understood by classical nucleation theory (CNT), which predicts the nucleation rate \cite{abraham1974homogeneous, debenedetti1996metastable, sear2014quantitative}
\begin{eqnarray}
R=N_S Zj{\rm e} \left( \frac{-\Delta G^*}{k_BT} \right),
\end{eqnarray}
where $\Delta G^*$ is the free energy barrier for forming a critical nucleus, which will be explicitly defined later. $k_BT$ represents the thermal energy with the Boltzmann constant $k_B$ and the absolute temperature $T$. $N_S$ is the number of nucleation sites. $j$ is the rate for monomers attaching to the nucleus. $Z$ is called the Zeldovich factor, which is the forward probability for a critical nucleus to grow diffusively into a larger nucleus rather than shrink back to nothing.

To further model the free energy barrier, CNT treats the microscopic nucleus as a macroscopic droplet, so that the free energy $\Delta G$ for forming a nucleus can be generally written as the sum of a bulk term proportional to the volume of the nucleus and a surface term proportional to its surface area. Especially for homogeneous nucleation, the nucleus modeled by a sphere of radius $r$ gives \cite{abraham1974homogeneous, debenedetti1996metastable, sear2014quantitative}
\begin{eqnarray}
\Delta G = \frac{4}{3} \pi r^3 \Delta g + 4 \pi r^2 \sigma.
\end{eqnarray}
The first term stands for the volume contribution, in which $\Delta g$ is the free energy difference per unit volume between the nucleated and non-nucleated phases and usually negative. The second term comes from the interface between the nucleus and its surroundings. $\sigma$ is the surface tension and always positive. As a consequence, for small $r$, the surface term dominates and $\Delta G(r)>0$; while for large $r$, the volume term dominates and $\Delta G(r)<0$. Especially at some intermediate value of $r$, the free energy goes through a maximum, corresponding to a least probability for nucleus occurring. This is called the critical nucleus and occurs at $dG/dr= 0$, which gives a critical nucleus radius
\begin{eqnarray}
r^* = -\frac{2 \sigma}{\Delta g}.
\end{eqnarray}
Adding new monomers to nuclei larger than this critical radius decreases the free energy, so the overall nucleation rate is then limited by the probability of forming the critical nucleus, which is $\Delta G^* = 16 \pi \sigma ^3/[3(\Delta g)^2]$. This is exactly the free energy barrier needed in the CNT expression for the nucleation rate $R$ above.

The reason for heterogeneous nucleation occurring much easier than homogeneous nucleation is that the nucleation barrier $\Delta G^*$ is much lower at a surface. For homogeneous nucleation, the nucleus is approximated by a sphere. However, for heterogeneous nucleation, when a nucleus is formed at the surface, its form is not completely spherical and depends on the contact angle \cite{fletcher1958size, hienola2007estimation}. This geometrical factor reduces the interfacial area and so the interfacial free energy, which in turn reduces the nucleation barrier. In principle, we expect nucleation to be fastest when the nucleus forms a small contact angle on its surface. However, detailed calculation is not straightforward and will not be listed here.

From the kinetic aspect, primary nucleation, including both both homogeneous and heterogeneous nucleation, is often modelled with the following formula \cite{tavare1995industrial, lomakin1997kinetic}
\begin{eqnarray}
\dfrac{dP}{dt}=k_n(m-m^*_{F})^{n_c},
\end{eqnarray}
where $k_n$ is the macroscopic reaction rate constant. With respect to the microscopic nucleation rate $R$ considered in above thermodynamic picture, we expect $\lim_{V\rightarrow\infty}R/N_s=k_n$ in the limit of sufficiently large volume. $m^*_{F}$ is the critical fibrillar concentration and $(m-m^*_F)$ is also known as supersaturation. Supersaturation is the driving force for both the initial nucleation step and the following growth, both of which could not occur in saturated or undersaturated conditions. $n_c$ stands for the critical nucleus size, that can be as large as 10, but generally ranges between 2 and 4.

In the 1960s, Oosawa \emph{et al.} borrowed this idea to study of the polymerisation of actin \cite{oosawa1962theory, oosawa1975thermodynamics} and wrote down a reaction scheme consisting of three basic process -- primary nucleation, monomer association and dissociation. However, they did not include the supersaturation effect explicitly, since mostly the protein concentration used for fibrillation is much higher than that required for saturation. According to Oosawa's model, the number concentration and mass concentration of aggregates evolve according to
\begin{eqnarray}
	&&\frac{d}{dt}P=k_nm^{n_c},\label{elongationP-2}\\
	&&\frac{d}{dt}M=k_e^+mP\underline{-k_e^-P+n_ck_nm^{n_c}}.\label{elongationM-2}
\end{eqnarray}
In above model, the inverse process for primary nucleation has been omitted. A major reason is that it is accounted by a boundary term $k_n^-[A_{n_c}]$ (where $k_n^-$ is the rate constant and $[A_{n_c}]$ is the concentration of critical nucleus), not compatible with other terms constituted by moments $M$ and $P$ (the moment-closure method we introduced in the section of how to coarse-grain model can systematically solve this problem). The last two terms underlying in (\ref{elongationM-2}) are generally negligible too, due to the fact that under proper fibrillation conditions, fiber elongation is more energetically favorable and proceeds faster than primary nucleation and monomer dissociation \cite{oosawa1975thermodynamics}. This leads to an important conclusion -- the major role of primary nucleation in fibrillation is to generate new seeds rather than directly consuming free monomers like elongation. Similar argument also applies to secondary nucleation introduced in the next section. Typical amyloid systems, which follow above mechanism without referring to secondary nucleation, include actins in KCl solvents \cite{wegner1982spontaneous} (see Fig. 9A), $\gamma$C-crystallin \cite{wang2010formation} and Apo C-II \cite{binger2008apolipoprotein} \textit{etc.}

According to \cite{cohen2011nucleated}, Oosawa's model admits analytical solutions as follows
\begin{eqnarray}
&&M(t) = m_{tot}-(m_{tot}-M_0)\big[\mu \textrm{sech}\big(\nu + \lambda \beta\mu t \big)\big]^{2/n_c},\label{Oosawa_result}\\
&&P(t)=P_0+k_n(m_{tot}-M_0)^{n_c}\mu(\beta\lambda)^{-1}\big[\tanh(\nu+\beta\lambda\mu t)-\tanh(\nu)\big],
\end{eqnarray}
where $\lambda = \sqrt{k_nk_e^+(m_{tot}-M_0)^{n_c} }$, $\beta =\sqrt{n_c/2}$, $\gamma = \beta k_e^+P_0/\lambda$, $\mu = \sqrt{1+\gamma^2}$ and $\nu = \textrm{arcsinh} (\gamma)$. $m_{tot}$, $M(0)=M_0$ and $P(0)=P_0$ are the total protein concentration, initial mass and number concentrations of aggregates respectively. Especially, in the absence of initial seeds $M_0=P_0=0$, we have $M(t) = m_{tot}[1-\textrm{sech}^{2/n_c}(\lambda\beta t)]$, which recovers the classical Oosawa's result \cite{oosawa1962theory}.

Insight into the early time behaviour of fiber mass concentration can be obtained by expanding (\ref{Oosawa_result}) around $t=0$
\begin{eqnarray}
M(t)=M_0+k_e^+P_0(m_{tot}-M_0)t+\frac{1}{2}(m_{tot}-M_0)[\lambda^2-(k_e^+P_0)^2]t^2+O(t^3),\quad t\rightarrow0.
\end{eqnarray}
This expression recovers the characteristic $t^2$ dependence relating to the primary nucleation in Oosawa's theory \cite{oosawa1962theory}. An additional term linear in time is raised by the growth of pre-added seeds. This difference actually provides a simple way to distinguish the seeding contributions from primary nucleation and pre-added seeds.

After some calculation, the half-time and the apparent fiber growth rate are given respectively as
\begin{eqnarray}
&&t_{1/2}=\frac{1}{\lambda\beta\mu}\big[\textrm{arccosh}(2^{n_c/2}\mu)-\nu\big],\\
&&k_{app}=\frac{\lambda\beta\mu}{n_c}\frac{m_{tot}+M_0}{m_{tot}}\sqrt{2^{n_c}\mu^2-1}.
\end{eqnarray}
According to above integrated rate laws, we can easily see that, in the absence of initial seeds $M_0=P_0=0$, the half time of fibrillation is proportional to $\lambda^{-1}$, which means $t_{1/2}\propto m_{tot}^{-n_c/2}$, a prominent feature of seeding through primary nucleation; on the contrary, if high concentrations of preformed seeds are added to the system, $\gamma\gg1$ and we recover results for pure elongation $t_{1/2}\propto(k_e^+P_0)^{-1}$.

Again, we can take advantage of this difference in scaling relations to separate the elongation process from primary nucleation. At first, without any initial seeds, the critical nucleus size for primary nucleation $n_c$ could be directly read out by examining the scaling dependence of half-time under different initial monomer concentrations. Furthermore, parameter $\lambda$ provides a combined knowledge of both primary nucleation and elongation. Finally, by adding high concentrations of performed seeds into the system, seeding through primary nucleation will be completely suppressed and the elongation rate could be extracted from the fibrillation kinetics solely. Similar argument is also applicable to secondary nucleation.

\begin{figure}[h]
	\begin{center}
		\includegraphics[width=0.7\textwidth,height=0.3\textwidth]{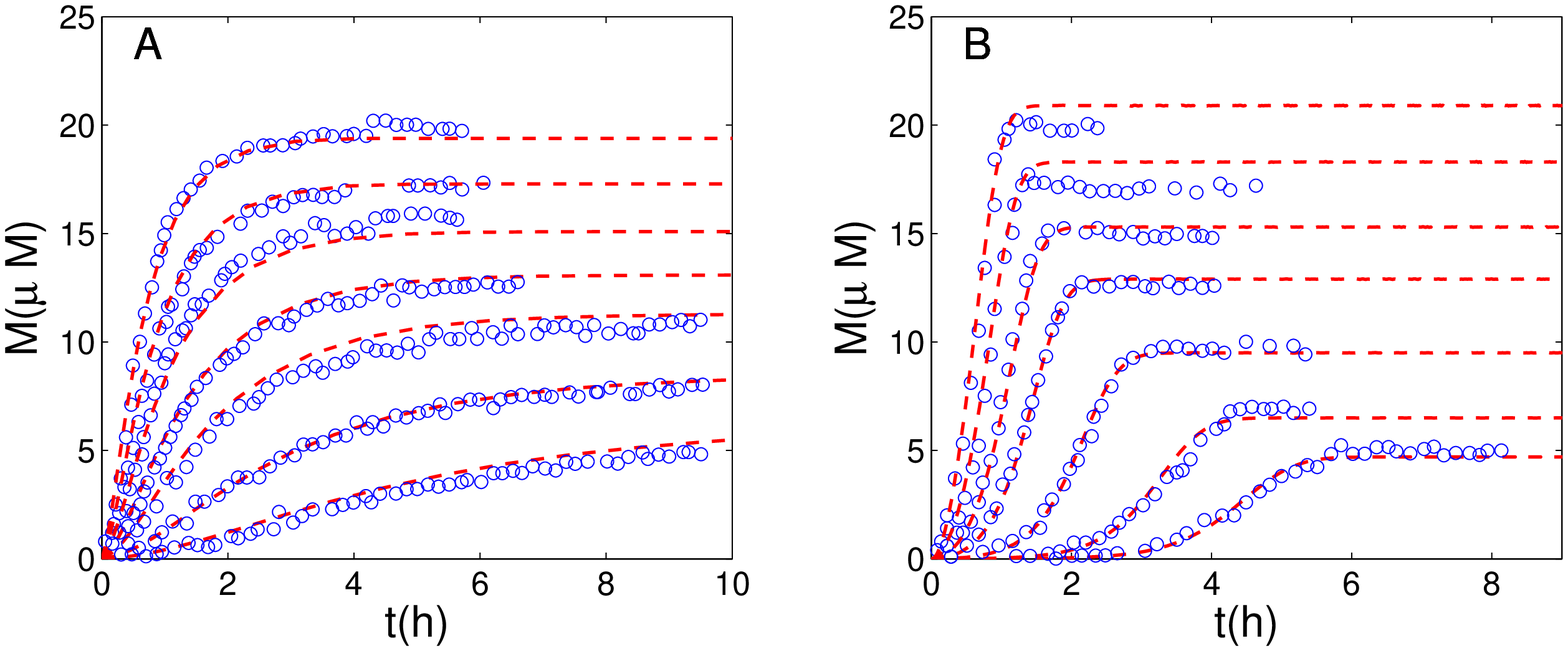}
       \caption{Actin growth in the presence of KCl v.s. MgCl$_2$, which highlights two different dominated mechanisms \cite{wegner1982fragmentation}. (A) Action samples were prepared with 40 mM KCl at monomer concentrations $m_{tot}$ = 7.4, 9.6, 12.4, 14.2, 16.2, 18.4, and 20.5 $\mu$M separately (blue circles). Predictions based on the NE model (red dashed lines) in (\ref{elongationP-2}) and (\ref{elongationM-2}) were performed with $n_c = 4$, $k_n=3\times10^{9}M^{-3}s^{-1}$,  $k_e^+=9\times10^2 M^{-1}s^{-1} $, $k_e^{-}=10^{-3}s^{-1}$. (B) Action samples were prepared with 0.6 mM MgCl$_2$ and 0.5 mM EGTA at monomer concentrations $m_{tot}$ = 6.7, 8.5, 11.5, 14.9, 17.3, 20.3, and 22.9 $\mu$M. Predictions based on the NEF model in (\ref{fragmentationP-1}) and (\ref{fragmentationM-1}) were performed with $n_c = 6$, $k_n=2\times10^{16}M^{-5}s^{-1}$,  $k_e^+=9\times10^4 M^{-1}s^{-1} $, $k_e^{-}=0.18s^{-1}$, $k_f^+=6\times10^{-7}s^{-1}$, $k_f^-=0$.}
		\label{Fig:NE model}
	\end{center}
\end{figure}

\subsubsection{Monomer-dependent secondary nucleation: self-catalysed process}
As indicated by the early time expansion of solutions for primary nucleation, a linear growth in the presence of initial seeds and a growth depending on $t^2$ for purely monomeric proteins are expected \cite{rochet2000amyloid}, which cannot account for the apparent high cooperativity of fibrillation observed in many amyloid systems. For example, in 1980 Ferrone \textit{et al.} \cite{ferrone1980kinetic} observed an exponential growth at the early stage of aggregation for the aberrant gelation of sickle-hemoglobin. Later, they studied the kinetics of polymerization of hemoglobin S and found a transition time much shorter than the lag phase predicted by primary nucleation only \cite{ferrone1985kinetics_1, ferrone1985kinetics_2}. Miranker \textit{et al.} observed a similar process in Islet amyloid polypeptide (IAPP) \cite{padrick2002islet}. Intriguingly, they found that both the reaction order and the activation enthalpy of two nucleation processes are identical, which made them to conclude that both primary nucleation and monomer-dependent secondary nucleation are ``alternative manifestations of the same, surface-catalyzed nucleation event'' \cite{ruschak2007fiber}. Above observations confirmed the existence of a secondary pathway for nucleation, which generates new nuclei by making use of the surface of existing aggregates in the system. Therefore, this kind of mechanism is called surface-catalysed secondary nucleation. Since it depends on both monomer concentration and fiber concentration, it is also known as monomer-dependent secondary nucleation, in contrast to a different kind of secondary nucleation mechanism -- fragmentation that is monomer independent. Just as Miranker claimed, monomer-dependent secondary nucleation is a special kind of heterogeneous nucleation and takes the surface of amyloid fibrils as catalyst. Therefore, it proceeds much faster than primary nucleation and exhibits an exponential growth due to the self-catalysed nature. Besides sickle-hemoglobin\cite{ferrone1980kinetic}, hemoglobin S \cite{ferrone1985kinetics_1, ferrone1985kinetics_2} and IAPP \cite{ruschak2007fiber}, A$\beta$40 \cite{meisl2014differences} and A$\beta$42 \cite{hellstrand2009amyloid, cohen2013proliferation} (See Fig. 10) have also been revealed to adopt the surface-catalyzed secondary nucleation mechanism.

A model, incorporating both primary nucleation and monomer-dependent secondary nucleation, could be expressed as
\begin{eqnarray}
	&&\frac{d}{dt}P=k_nm^{n_c} + k_2m^{n_2}M,\label{secondaryP-1}\\
	&&\frac{d}{dt}M=k_e^+mP-k_e^-P \underline{+n_ck_nm^{n_c} + n_2k_2m^{n_2}M},\label{secondaryM-1}
\end{eqnarray}
where the new term $k_2m^{n_2}M$ accounts for the contribution of secondary nucleation on seeding, which is proportional to the surface area of existing aggregates (scale as $M$). The parameter $n_2$ stands for the critical nucleus size for secondary nucleation, analogous to $n_c$ for primary nucleation. It is noted that when $n_2=0$, above equations offers a good approximation to the model of fragmentation, an alternative secondary nucleation mechanism going to be shown in the next section. Therefore, we can use the same model to describe both monomer dependent and independent secondary nucleations by just tuning the parameter $n_2$. As we have stated in the section of primary nucleation that the major role of various nucleation processes is to create new sites for elongation rather than increasing the mass of aggregates directly, the terms underlying in (\ref{secondaryM-1}) could be neglected with respect to elongation.

Under this condition, first-order self-iterative solutions have been obtained through fixed-point analysis \cite{granas2013fixed} by Cohen \emph{et al.} \cite{cohen2011nucleated_2} as
\begin{eqnarray}
&&M(t)=M(\infty)-[M(\infty)-M_0]e^{-k_{\infty}t}\bigg(\frac{B_-+C_+e^{\kappa t}}{B_++C_+e^{\kappa t}}\cdot\frac{B_++C_+}{B_-+C_+}\bigg)^{k_{\infty}^2/(\kappa \bar{k}_{\infty})},\label{approximateM-1}\\
&&P(t)=\frac{P_l(t)}{1+P_l(t)/P(\infty)}\label{approximateP-1},
\end{eqnarray}
in which $P_l(t)=(C_+\kappa e^{\kappa t}+C_-\kappa e^{-\kappa t})/k_e^+$ is the linearized solution for early time. $\kappa=\sqrt{(k_e^+m_0-k_e^-)k_2m_0^{n_2}}$, $\lambda=\sqrt{k_e^+k_nm_0^{n_c}}$, $C_{\pm}=k_e^+P_0/(2\kappa)\pm k_e^+M_0/[2(m_0k_e^+-k_e^-)]\pm \lambda^2/(2\kappa^2)$, $k_{\infty}=k_e^+P(\infty)$, $\bar{k}_{\infty}=\sqrt{k_{\infty}^2-4C_+C_-\kappa^2}$, $B_{\pm}=(k_{\infty}\pm\bar{k}_{\infty})/(2\kappa)$. $M(0)=M_0$, $P(0)=P_0$ and
$M(\infty)=m_{tot}-k_e^-/k_e^+$, $P(\infty)=(k_e^+)^{-1}\sqrt{2\kappa^2/[n_2(n_2+1)]+2\lambda^2/n_c+2M_0\kappa^2/(n_cm_0)+(k_e^+P_0)^2}$ are the aggregates number concentration and mass concentration at the beginning and in the equilibrium respectively.  $m(0)=m_0=m_{tot}-M_0$ and $m_{tot}$ are the initial and total monomer concentrations. Alternative solutions for early time based on perturbation methods could be found in \cite{ferrone1985kinetics_2, ferrone199917} and we will not go into them here.

According to self-iterative solutions, the half-time and apparent fiber growth rate can be extracted,
\begin{eqnarray}
&&t_{1/2}\approx \kappa^{-1}\ln(1/C_+),\label{approximate-t}\\
&&k_{app}\approx\frac{\kappa}{2}\frac{k_{\infty}^2/(\kappa \bar{k}_{\infty})}{B_++1}\label{approximate-k}.
\end{eqnarray}
It is clear, in the absence of initial seeds $M_0=P_0=0$, the half time of fibrillation is around $2\kappa^{-1}\ln(\kappa/\lambda)$, which means $t_{1/2}\propto m_{tot}^{-(n_2+1)/2}$, a feature of seeding through secondary nucleation. As primary nucleation only enters into the half-time through a logarithmic correction, its influence will be limited to early time, which is expectable since secondary nucleation is more effective in generating seeds. Furthermore, even if medium amount of preformed seeds are introduced to the system $\lambda\ll k_e^+P_0\ll\kappa$, the scaling dependence of half-time on monomer concentration will not be changed except for a logarithmic correction, showing only primary nucleation is screened out in this step. Unless high concentrations of preformed seeds are added $k_e^+P_0\gg\kappa$, we can not eliminate the effect of secondary nucleation and recover the result for pure elongation $t_{1/2}\propto(k_e^+P_0)^{-1}$.

Since the major role of primary nucleation and secondary nucleation is to generate new seeds, it is generally expected, by introducing preformed seeded into the system, various nucleation mechanisms could be partially or even completely screened out. The examination above confirms this point from an analytical point. In particular, we see that firstly primary nucleation and then secondary nucleation are suppressed with an increase of seeds concentration. As long as two nucleation processes are well separated in time, given by $\sqrt{k_e^+k_nm_{tot}^{n_c}}\ll\sqrt{k_e^+k_2m_{tot}^{n_2+1}}$, rates of primary nucleation, secondary nucleation and elongation could be extracted respectively by simply varying the seeds concentration. In this way, the significant role of seeds in fibrillation kinetics could be fully appreciated. All these discussions are applicable to the fragmentation case in the next section by just setting $n_2=0$.

\begin{figure}[h]
	\centering
		\includegraphics[width=1\textwidth]{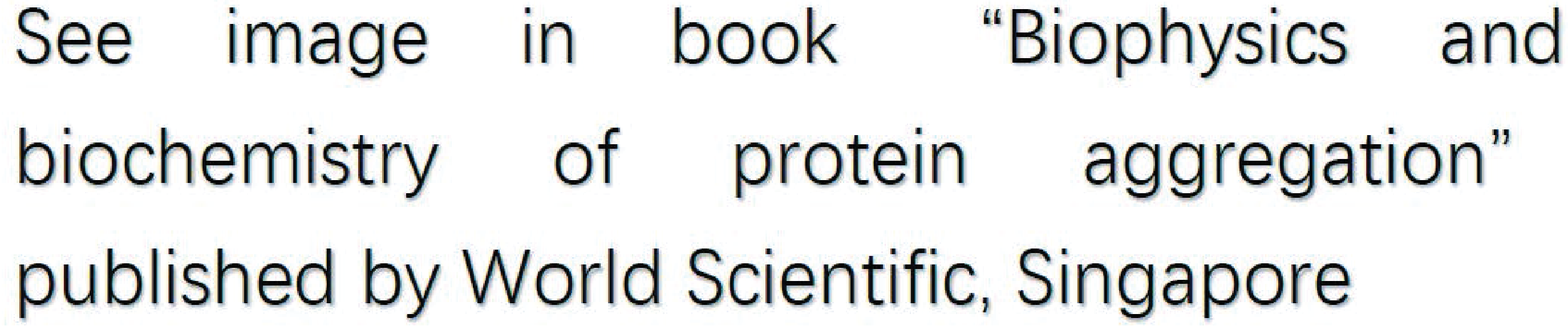}
		\caption{Kinetics of A¦Â42 aggregation is shifted from surface catalyzed secondary nucleation dominated mechanism to fragmentation dominated by continuously increasing agitating speeds. The upper plots show the time profile of A¦Â42 aggregation under different shear rates generated by agitating the sample under different speeds; and the lower ones show the corresponding power-law relationships between the half-time and the initial monomer concentration of A¦Â42. In (A), the rate parameters $\sqrt{k_e^+k_n}=42.4M^{-1}s^{-1}$ and $\sqrt{k_e^+k_2}=2.8\times 10^5M^{-3/2}s^{-1}$, $n_c=n_2=2$ are fixed and $k_f^+ = k_f^-=0 $. In (B)-(F), $\sqrt{k_e^+k_f^+}=0.6, 0.9, 1.4, 1.9M^{-1/2}s^{-1}$ are taken respectively under each shear rate in replace of $\sqrt{k_e^+k_2}$. Figure is taken from \cite{cohen2013proliferation} with copyright permission.}
		\label{Fig:NEF model}
\end{figure}

\subsubsection{Monomer-independent secondary nucleation: fragmentation and annealing}
Another efficient way to generate seeds without involving primary nucleation is fragmentation, which means one filament breaks into two smaller fragments. In principle, breaking into three or more pieces simultaneously is also possible but extremely rare. The universality of fragmentation in amyloid formation has long been established in the famous book of Oosawa \cite{oosawa1975thermodynamics}. It is well-known that single filament becomes mechanically unstable and tends to break when exceeding certain threshold in length \cite{collins2004mechanism,knowles2009analytical}. Even for bundled fibrils, breakage is unavoidable in the presence of mechanical stress {\cite{sanders1995skin}}, thermal motion  \cite{lee_1dbreakage, loveday2012beta, lee2015thermal}, or chaperons like Hsp104 {\cite{tessarz2008substrate}}.

Compared to surface-catalysed secondary nucleation, fragmentation is monomer-independent and thus becomes predominant under the condition of low monomer concentration. It can dramatically accelerate the formation of breakable filaments and change time courses for mass concentration of aggregates into a sigmoidal like behavior. Fragmentation could further alter the fiber length distribution globally just like elongation and enhance the toxicity of fibril samples by generating more harmful oligomers in low molecular weight. Due to its significant roles in both amyloid fibrillation and cytotoxicity, fragmentation has been paid plenty of attention to in past studies. And many typical amyloid systems have been shown to follow this mechanism, including actins in MgCl$_2$ solvents \cite{wegner1982spontaneous} (see Fig. 9B), Sup35 NW region \cite{collins2004mechanism}, Ure2p \cite{zhu2003relationship}, $CsgB_{trunc}$ \cite{hammer2007curli}, Stefin B \cite{vskerget2009mechanism}, $\beta$2-microglobulin \cite{xue2008systematic}, WW domain \cite{ferguson2003rapid}, $\alpha$-synucleins \cite{uversky2001metal} and insulin \cite{nielsen2001effect} \textit{etc.}

Generally speaking, fragmentation rates are length-dependent. In literature, several assumptions have been proposed, including the random scission, the central scission, and the Gaussian scission \cite{ballauff1981degradation,ballauff1984degradation} and modified partially random scission \cite{tanigawa1996changes} \textit{etc.} The latter two got supports from fiberlike PI264-b-PFS48 micelles under sonication \cite{guerin2008fragmentation}, thermodynamically induced shear degradation of polystyrene in semiconcentrated solutions \cite{ballauff1984degradation}, single-stranded random-coiled poly-(uridylic acid), double-stranded helical DNA, and triple-stranded helical poly-(adenylic acid)$\cdot$2poly(inosinic acid) \cite{tanigawa1996changes}. Hill suggested a formula in log form for polymer fragmentation and annealing based on statistical mechanics \cite{hill1983length}. And it has been applied by Hong \textit{et al.} to the study of several amyloid proteins \cite{hong2013simple}.

In the presence of fiber-length-dependent fragmentation, the derivation of self-closed models for fiber formation becomes a big trouble. Based on Maximum Entropy Principle \cite{jaynes1957information,jaynes1982rationale,presse2013principles}, a systematical and reliable way has been proposed to derive close-formed mass-action equations from microscopic length-dependent fragmentation models \cite{hong2013simple} (see section on how to coarse-grain model for details). In particular, under the assumption of totally random scission (or length-independent fragmentation), the kinetics of fiber formation can be expressed by the following model:
\begin{eqnarray}
&&\frac{d}{dt}P=k_nm^{n_c} + k_f^+[M-(2n_c-1)P]-k_f^-P^2,\label{fragmentationP-1}\\
&&\frac{d}{dt}M=k_e^+mP - k_e^-P\underline{+n_ck_nm^{n_c}},\label{fragmentationM-1}
\end{eqnarray}
where $k_f^+$ is the rate constant for length-independent fragmentation, and $k_f^-$ is that for the inverse process -- filaments annealing. Terms for fiber fragmentation and annealing in (\ref{fragmentationP-1}) are explicit and can be obtained directly from the microscopic model. Since here fibrils are not allowed to break into monomers directly and so is annealing, no term for fragmentation or annealing will enter into the equation for $M(t)$. This is a dramatic feature of fiber fragmentation and annealing, which only affect the number concentration of aggregates and keep the mass concentration conserved.

As an inverse process of fragmentation, filaments annealing is crucial for the fiber length distribution, number concentration of aggregates as well as their average length. Based on (\ref{fragmentationP-1}), it is clearly seen that in the absence of annealing (by setting $k_f^-=0$), the average length of fibrils will be around $2n_c-1$, in contrast to the estimated lower bounds of at least hundreds of monomers.

Again, by fixed-point analysis, self-iterative solutions could be derived in exactly the same form as those in (\ref{approximateM-1}) and (\ref{approximateP-1}), except for some internal parameters have to be changed correspondingly \cite{michaels2014role}, \textit{i.e.} the linearized solution for early time $P_l(t)=C_1e^{k_1t}+C_2e^{k_2t}-n_ck_f^+\epsilon$,
number concentration of aggregates in equilibrium $P(\infty)=2M(\infty)\big(2n_c-1+\sqrt{(2n_c-1)^2+4k_f^-M_0/k_f^+}\big)^{-1}$, $\kappa=\sqrt{(k_e^+m_0-k_e^-)k_f^+}$, as well as $C_{\pm}=k_e^+C_{1,2}/k_{1,2}$, where $k_{1,2}=-k_f^-P_0\pm\sqrt{(k_f^-P_0)^2+\kappa^2}$, $C_{1,2}=(1-k_{2,1}/k_{1,2})^{-1}[P_0-k_{2,1}(M_0+k_nm_0^{n_c}/\kappa^2+n_ck_f^+k_f^-P_0\epsilon/\kappa^2)]$. The same expressions for the half-time and apparent fiber growth rate could be obtained as in (\ref{approximate-t}) and (\ref{approximate-k}) and will not be addressed here. Just note in current case, except for a logarithmic correction, $t_{1/2}\propto m_{tot}^{-1/2}$ and $k_{app}\propto m_{tot}^{1/2}$, a special case of monomer-dependent secondary nucleation with $n_2=0$ for fragmentation as we claimed.

Models constituted by above four basic mechanisms: elongation, primary nucleation, surface-catalyzed secondary nucleation and fragmentation (as well as monomer dissociation and filaments annealing as two typical inverse processes) could already explain most fibrillation kinetics observed in experiments. In the past several years, fruitful results have been obtained in this direction \cite{cohen2012macroscopic,gillam2013modelling}, which greatly enhance our understandings on the underlying microscopic processes, the kinetics and thermodynamics of amyloid fibrillation, effects of pH value, temperature and \textit{etc.} Besides this main framework, several key issues need to be addressed to provide a complete picture. Firstly, we have mentioned that high monomer concentration could dramatically change the kinetics of amyloid formation, which is known as saturation. Luckily, the effect of saturation could be easily accounted through a Michaels-Menton-like formula, though model analysis becomes far more difficult. Secondly, it is well-known that the formation of amyloid fibrils is a direct consequence of protein misfolding, and the conformational conversion of monomers and oligomers plays an inreplaceable role in it. However, the step of conformational conversion is easily neglected in kinetics due to the interference of primary nucleation.

\subsection{How high concentration affect}
In the section of fiber growth, we have mentioned that fiber elongation shows a sub-linear dependence in the regime of high monomer concentration. In fact, this phenomenon is rather universal and has a deep physical basis on saturation. It is imaginable, in the presence of too many monomers competing for the same fibril end at the same time, the fiber end will appear to be ``saturated'' since the incorporation of each monomer requires certain amount of time and can not be finished at once. As a consequence, the elongation process appears to be blind to the instantaneous monomer concentration in the system and shows a weaker-than-linear dependence.

Physically, the process of saturated elongation could be modeled through a ``dock-lock'' mechanism \cite{esler2000alzheimer,nguyen2007monomer}, which includes two sub-steps -- unspecific attachment and detachment of a monomer with the fibril end, and subsequent conformational change of the attached monomer to make the attachment specifically. Usually, the second process is the rate-limiting step. For example, Scheibel \textit{et al.} studied how nuclei mediate the conversion of soluble NM domain of Sup35 to the amyloid form in the elongation phase of fiber formation \cite{scheibel2004elongation}. By creating single-cysteine substitution mutants at different positions of NM domain to provide unique attachment sites for various probes, they established that elongation is a two-step process involving the capture of an intermediate, followed by its conformational conversion.

If we take $P_{bound}$ and $P_{free}=P-P_{bound}$ as the number concentration of fibril ends which has and has not monomers unspecifically attached, the ``dock-lock'' mechanism is expressed through following equations
\begin{eqnarray}
&&\frac{d}{dt}M=k_cP_{bound},\label{saturationM-1}\\
&&\frac{d}{dt}P_{bound}=k_a^+mP_{free}-k_a^-P_{bound}-k_cP_{bound},\label{saturationP-1}
\end{eqnarray}
where $k_a^+$, $k_a^-$ and $k_c$ represent reaction rate constants for monomer unspecific attachment, detachment and conformational change at the fibril end. It is clear that the three terms on the right-hand side of (\ref{saturationP-1}) account for the contribution of each step mentioned in the ``dock-lock'' mechanism to fibril ends respectively.

To eliminate the additional variable $P_{bound}$, we refer to the classical Quasi Stead-State Approximation \cite{bodenstein1924quasi,benson1960foundations}, which assumes the generation and consumption of $P_{bound}$ are always in a dynamical equilibrium. Such that we can take the sum of terms on the right-hand side of (\ref{saturationP-1}) to be zero, \textit{i.e.}
\begin{eqnarray}
0=k_a^+mP_{free}-k_a^-P_{bound}-k_cP_{bound},\nonumber
\end{eqnarray}
which gives solutions in a form of the famous Michaelis-Menten equation for enzyme kinetics \cite{michaelis1913kinetik},
\begin{eqnarray}
&&P_{free}=P\bigg(1+\frac{m}{K_m}\bigg)^{-1},\nonumber\\
&&P_{bound}=P\frac{m}{K_m}\bigg(1+\frac{m}{K_m}\bigg)^{-1}.\nonumber
\end{eqnarray}
Here the Michaelis constant $K_m=(k_a^-+k_c)/k_a^+$.

Inserting above formula into (\ref{saturationM-1}), a simplified model incorporating the process of saturated elongation is reached,
\begin{eqnarray}
&&\frac{d}{dt}M=k_e^+m P\bigg(1+\frac{m}{K_m}\bigg)^{-1},\\
&&\frac{d}{dt}P=k_nm^{n_c}+k_2m^{n_2}M,
\end{eqnarray}
in which the new rate constant for fiber elongation is defined as $k_e^+=k_c/K_m=k_c k_a^+/(k_a^-+k_c)$. Compared to the formula in the section of fiber growth, a correction factor $(1+m/K_m)^{-1}$ has been added. Therefore, by introducing an ``effective'' monomer concentration as $m/(1+m/K_m)$, above model will recover the classical one without saturation. Furthermore, if monomer concentration is much higher than the critical saturation concentration (given by the Michaelis constant $K_m$) $m\gg K_m$, the effective monomer concentration becomes a constant $K_m$; otherwise if the monomer concentration is much lower than the critical saturation concentration $m\ll K_m$, the effective monomer concentration approaches to the real monomer concentration. In this way, both the linear dependence of fiber elongation rate in the regime of low monomer concentration and sub-linear dependence in the high regime could be explained by a unified picture based on saturated elongation. And the Michaelis constant $K_m$ serves as a key to characterize the transition from linear to sub-linear (Fig. 11A and 11B).

The model for saturated elongation is far more difficult to solve than the classical one without saturation. In the absence of initial seeds, a suggested practical solution is (unpublished result)
\begin{eqnarray}
M(t)=m_{tot}\bigg\{1-\bigg[1+\frac{1}{\theta}y(1+y)^{\alpha/(1+\alpha)}\bigg]^{-\theta}\bigg\},
\end{eqnarray}
where $y=\epsilon e^{\kappa t/\sqrt{1+\alpha}}$, $\alpha=m_{tot}/K_m$, $\theta=\sqrt{2/[n_2(n_2+1)]}$ and $\kappa=\sqrt{k_e^+k_2m_{tot}^{n_2+1}}$. A critical concentration of fibrils $M(0)/m_{tot}=\epsilon=k_n^+m_{tot}^{n_c-n_2-1}/(2k_2)\ll1$ has been introduced to seed the system, so that the resulting expression for $M(t)$  matches the leading order term for early time. In the special case of $n_2=0$, which corresponds to fragmentation, the solution reduces to
\begin{eqnarray}
M(t)=m_{tot}\bigg\{1-{\rm e}\bigg[-y(1+y)^{\alpha/(1+\alpha)}\bigg]\bigg\},
\end{eqnarray}
by exploiting the identity $\lim_{b\rightarrow\infty}(1+a/b)^b = e^a$. Meanwhile, we have
\begin{eqnarray}
P(t)=\frac{\sqrt{\theta^2+\alpha\vartheta^2}\kappa}{2k_+}\bigg[1-\bigg(\frac{m}{m_{tot}}\bigg)^{1/\theta}\bigg],
\end{eqnarray}
where $\vartheta=\sqrt{2/[(n_2+1)(n_2+2)]}$. Again, we need to pay attention to the case $n_2=0$, which gives
\begin{eqnarray}
P(t)=\frac{1}{2}k_n m_{tot}^{n_c}t+\frac{k_2m_{tot}}{\kappa}\ln\bigg(\frac{1+y}{1+\epsilon}\bigg).
\end{eqnarray}

Based on above solutions, we can roughly determine the half time and the apparent fiber growth rate as
\begin{eqnarray}
&&t_{1/2}\sim\ln(1/\epsilon)\sqrt{1+\alpha}/\kappa,\\
&&k_{app}\sim\kappa/\sqrt{1+\alpha}.
\end{eqnarray}
It is easily seen that $t_{1/2}\propto m_{tot}^{-(n_2+1)/2}$ in the regime of low monomer concentrations $m_{tot}\ll K_m$ and $t_{1/2}\propto m_{tot}^{-n_2/2}$ in the regime of high monomer concentrations $m_{tot}\gg K_m$. Especially, when $n_2=0$ and $m_{tot}\gg K_m$, we have $t_{1/2}\propto ln(1/m_{tot})$, which well explains the observed sub-linear dependence of fiber elongation under the condition of high monomer concentrations.

\begin{figure}[h]
	\centering
		\includegraphics[width=0.7\textwidth]{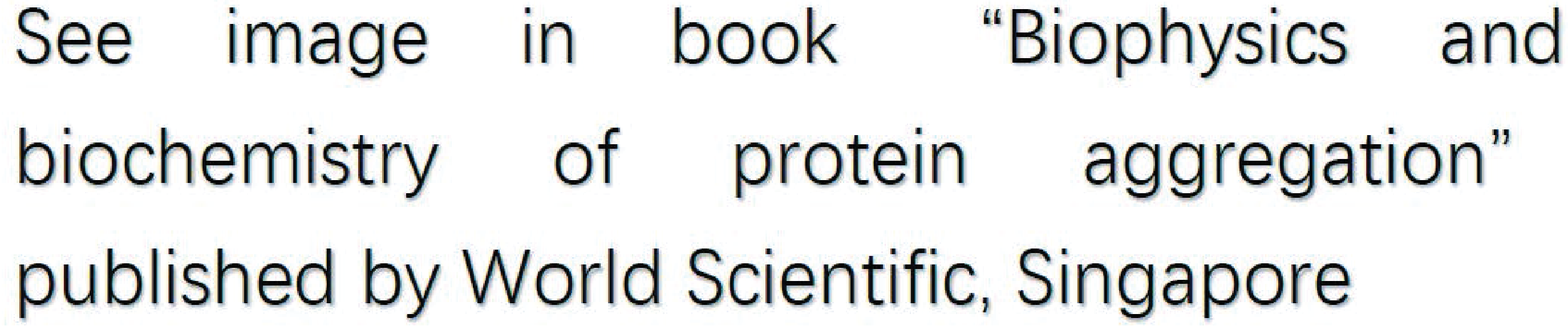}
		\caption{(A)-(B): Saturation of elongation rate for $\alpha$-synuclein under different concentrations
			of soluble proteins and a constant $3.5\mu M$ seeds. Images are taken from
			\cite{buell2014solution}. (C)-(D): Saturation of surface catalyzed secondary nucleation for A$\beta$40 under various monomer concentrations without initial seeds. Data fitting was performed according to (\ref{saturationP-2}) and (\ref{saturationM-2}) with $k_e^+=6\times10^5M^{-1}s^{-1}$, $k_n=2\times10^{-6}M^{-1}s^{-1}$, $k_2=3\times10^3M^{-2}s^{-1}$, $K_s=6\times10^{-6}M$, $n_c=n_2=2$. Images are taken from \cite{meisl2014differences} with copyright permission.}
		\label{Fig:NEF model}
\end{figure}

In principle, all monomer-dependent processes may get saturated once the monomer concentration exceeds certain threshold. And large amyloid proteins are more prone to get saturated than smaller ones under the same condition, since the former generally requires a longer time to fit itself to the fibrillar structure. In view of saturation, the model for surface catalyzed secondary nucleation (\ref{secondaryP-1}) requires to be modified too (see Fig. 11C and 11D). Following a similar ``dock-lock'' mechanism as well as derivations for saturated elongation, saturated secondary nucleation could be formulated as
\begin{eqnarray}
&&\frac{d}{dt}P=k_nm^{n_c}+k_2\frac{m^{n_2}}{1+(m/K_s)^{n_2}}M,\label{saturationP-2}\\
&&\frac{d}{dt}M=k_e^+mP\label{saturationM-2},
\end{eqnarray}
where $K_s$ is the critical saturation concentration for secondary nucleation. Again, if monomer concentration is much lower than the critical saturation concentration $m\ll K_s$, we will recover the classical model without saturation; contrarily, if monomer concentration is much higher $m\gg K_s$, only a constant concentration of monomers $K_s$ could contribute to secondary nucleation due to saturation.

Self-iterative solutions in a similar form of (\ref{approximateM-1}) and (\ref{approximateP-1}) could be derived for above model through fixed-point analysis. Interested readers may refer to \cite{meisl2014differences} for details. Basically, everything is the same except $k_2$ is replaced by $k_2/[1+(m_{tot}/K_s)^{n_2}]$. Of particular interest is the scaling behavior of half-time, which are solved within logarithmic corrections as $t_{1/2}\approx\kappa^{-1}\ln(1/C_+)$, where $\kappa=\sqrt{k_e^+k_2m_0^{n_2+1}/[1+(m_0/K_s)^{n_2}]}$, $C_+=k_e^+P_0/(2\kappa)+k_e^+M_0/(2m_0k_e^+)+\lambda^2/(2\kappa^2)$, $\lambda=\sqrt{k_e^+k_nm_0^{n_c}}$. And $m(0)=m_0$, $P(0)=P_0$, $M(0)=M_0$ are the monomer concentration, number and mass concentration of seeds at the start of reaction respectively. Thus, in the presence of high monomer concentration $m_0\gg K_s$, we have $t_{1/2}\propto m_{tot}^{-1/2}$, similar as that for the model of fragmentation; while in the regime of low monomer concentration $m_0\ll K_s$, the half-time $t_{1/2}\propto m_{tot}^{-(n_2+1)/2}$, which recovers the unsaturated case as expected. The model combined both saturated elongation and saturated secondary nucleation is similar and will not be shown here.

\subsection{How to incorporate different conformations}
It is well-known that the formation of amyloid fibrils is a direct consequence of the misfolding of amyloid proteins \cite{sunde1998globular, dobson2003protein}. As an example, in one of the most well studied amyloid systems -- prion, Prusiner identified that the conformational conversion of prion protein from PrP to PrP$^{sc}$ gives rise to the famous mad cow disease and scrapie in sheep \cite{prusiner1982novel}. In human, prion causes Creutzfeldt-Jakob disease (CJD) and kuru. According to the molecular size, conformational conversion could be divided into either monomer conversion or oligomer conversion. While based on the position where the conformational conversion happens inside the whole fibrillation procedure, it can be classified into pre-, on-pathway and off-pathway conversion separately.

Among them, pre-conversion is a prerequisite for primary nucleation, elongation and surface catalyzed secondary nucleation. It is directly related to monomer conversion and means the conformation of monomers has to be adjusted in order to be incorporated into fibrillar structures \cite{kelly1998alternative, xing2002induction}. The kinetic modeling of conformational pre-conversion is most straightforward. All terms concerning instantaneous monomer concentration $m(t)$ in previous models should be replaced by a new concentration of monomers in the unfolded (or partially unfolded) state $m_u(t)$ as required by the fibrillar structure \cite{hong2011dissecting}.
\begin{eqnarray}
m_u(t)=\int_0^tk_c^+m(\tau)e^{-(k_c^++k_c^-)(t-\tau)}d\tau+m_u(0)e^{-(k_c^++k_c^-)t},
\end{eqnarray}
is the solution of
\begin{eqnarray}
\frac{dm_u}{dt}=k_c^+(m-m_u)-k_c^-m_u,\label{conversion}
\end{eqnarray}
where $k_c^+$ and $k_c^-$ are the forward and backward reaction rate constants for protein conversion between the folded and unfolded states. In practice, the effect of conformational pre-conversion is easily neglected due to the existence of other rate-limiting steps, like primary nucleation. This fact is fully appreciated based on following argument. If conformational conversion of monomers is faster than fiber growth rate, which is generally limited by primary nucleation, we could assume monomers in the folded and unfolded states are in dynamic equilibrium and apply QSSA approximation to obtain $k_c^+(m-m_u)-k_c^-m_u=0$. By redefining $\tilde{k}_n=[k_c^+/(k_c^++k_c^-)]^{n_c}k_n$, $\tilde{k}_e^+=[k_c^+/(k_c^++k_c^-)]k_e^+$, $\tilde{k}_2=[k_c^+/(k_c^++k_c^-)]^{n_2}k_2$, the model in (\ref{secondaryP-1}) and (\ref{secondaryM-1}) is recovered without including conformational pre-conversion explicitly.

In contrast to pre-conversion, objects of on-pathway and off-pathway conversion are both oligomers. Their main difference lies on whether conformational conversion is necessary for fibril generation or not. In the on-pathway conversion, oligomers need to take some rearrangements in structure, for example repacking from globular oligomeric conformation to linear fibril-like conformation, before growing into mature fibrils; while in the off-pathway conversion, conformational conversion of oligomers will leads to amorphous aggregates in competition with fibrillar aggregates. The kinetic modeling of on-pathway and off-pathway conversions is far more difficult than pre-conversion, due to the conformational variety of oligomers, complicated interactions between oligomers, fibrils and amorphous aggregates \textit{etc.} Preliminary results for on-pathway conversion leading to fibrillary aggregates in transthyretin(TTR) \cite{peterson1998inhibiting}, tau proteins \cite{shammas2015mechanistic}, insulin stabilized by Zn$^{2+}$ \cite{lee2007three} \textit{etc.}, and off-pathway conversion leading to the formation of amorphous aggregates \cite{kodaka2004interpretation, kodali2007polymorphism, powers2008mechanisms} could be found in literature. But details will be omitted with a pity.

Besides those well-formulated mechanisms discussed in current paper, there are still many processes, which play an important role in changing the morphology of fibrils, affecting fibril thermodynamic stability, modifying fibrillation kinetic profiles \textit{etc.}, worthy of exploring. For instance, Anderson \textit{et al.} observed glucagon fibrils are able to generate new fibril ends by continuously branching, which prefers an angle of $35^o-40^o$ along the forward direction of parent fibril and never occurs at the tip \cite{andersen2009branching}. Murphy \textit{et al.} further included lateral aggregation of filaments into fibrils to build up a complete description of A$\beta$ fibrillation \cite{pallitto2001mathematical}. Most importantly, in vivo conditions are completely different from that in vitro \cite{dobson2004principles, luheshi2008protein}. Not only the cellular crowding environment \cite{munishkina2004effect, magno2010crowding}, but also synthesis \cite{hung1993activation}, degradation \cite{wang2006beta} and transportation \cite{buxbaum1998alzheimer, gunawardena2001disruption} of amyloid proteins may exert a great impact on the fibrillation kinetics. However, currently we still lack a quantitative characterization for most of them. These interesting topics have to be left to the future with regrets.

\begin{figure}[h]
	\centering
		\includegraphics[width=1\textwidth,height=0.39\textwidth]{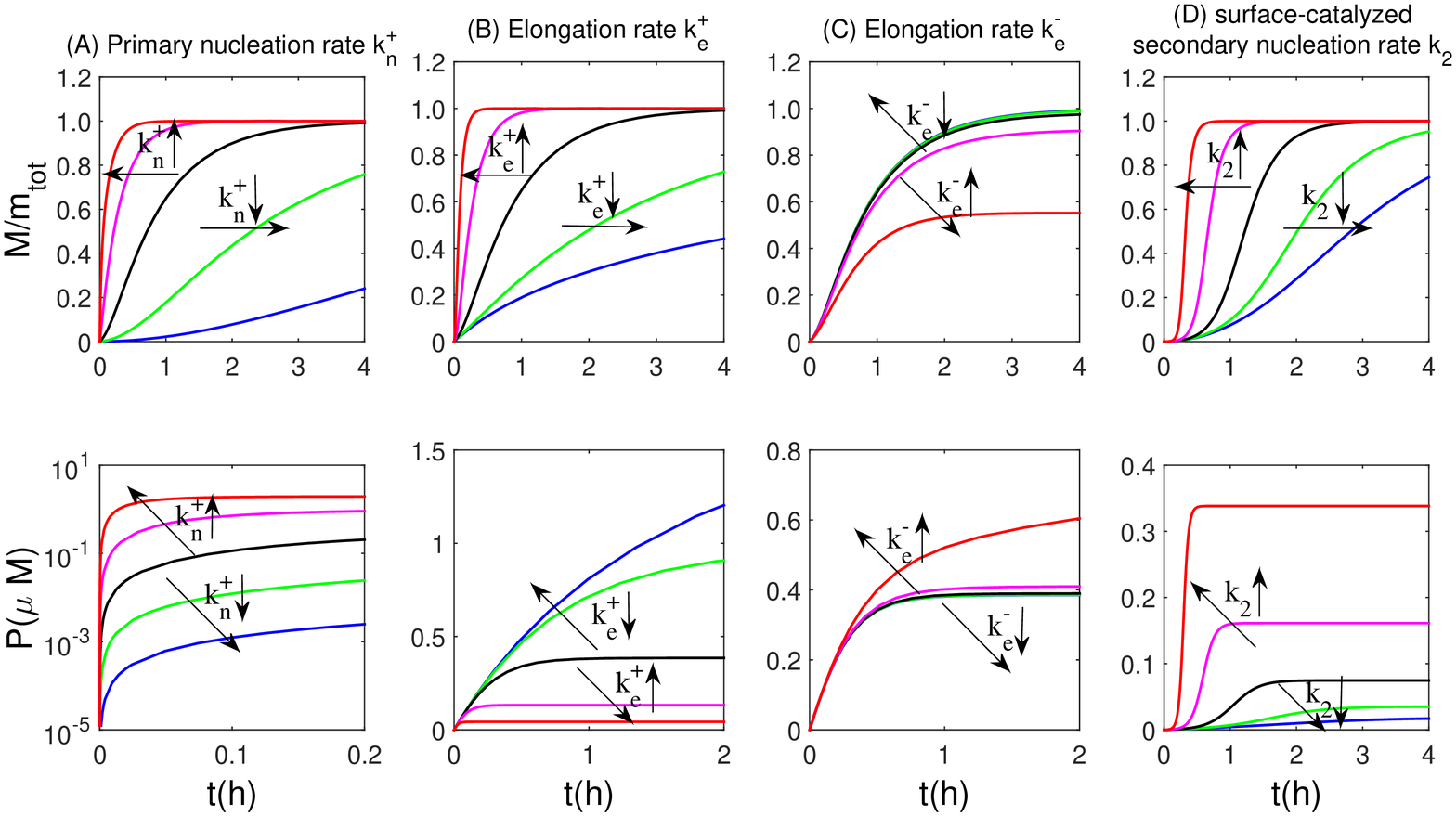}
        \includegraphics[width=1\textwidth,height=0.39\textwidth]{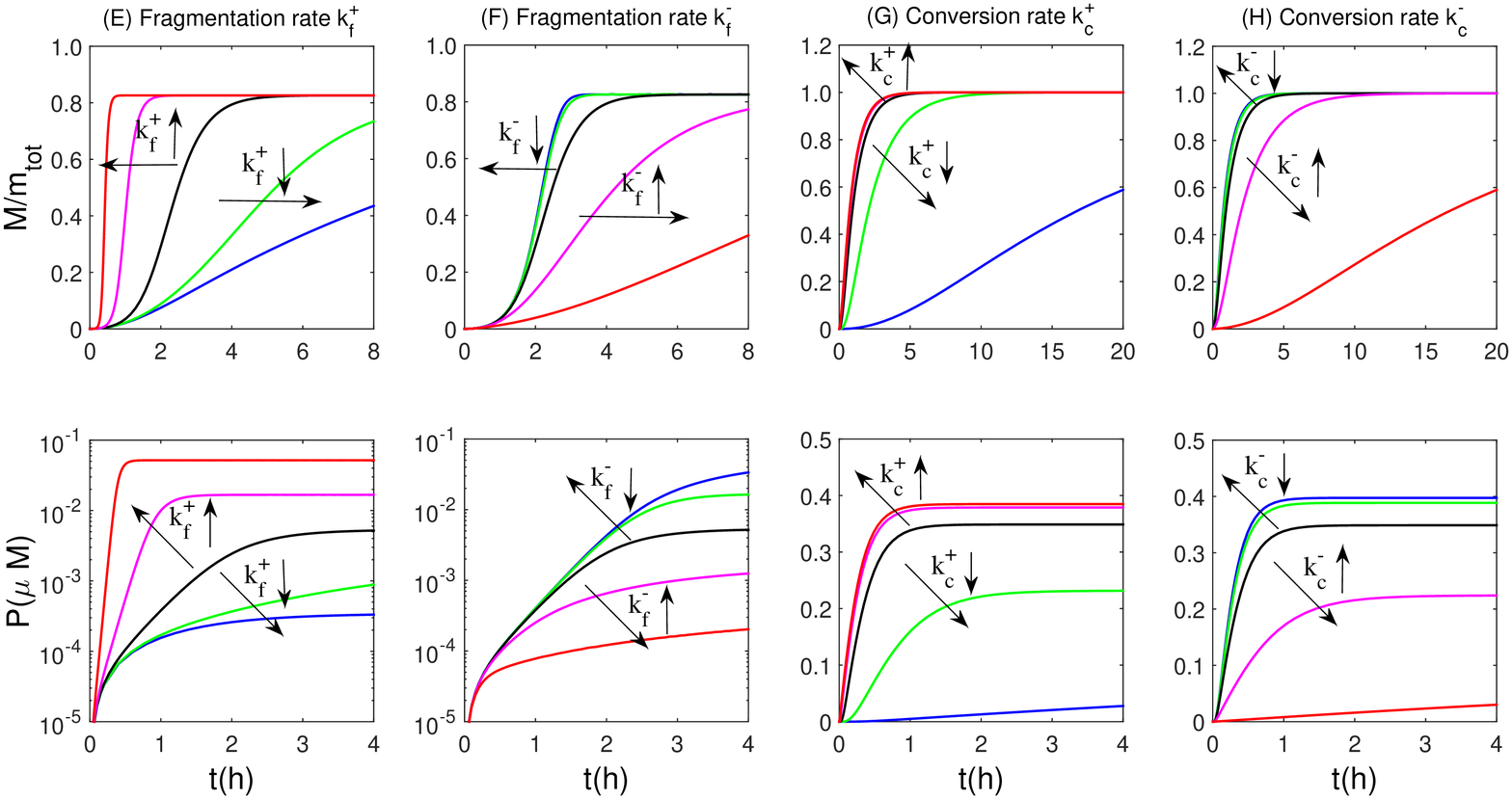}
		\caption{Influence of rate constants on the fibrillation kinetics. In (A-C), rates for primary nucleation, elongation and monomer dissociation are changed with respect to those in Fig. 9A ($m_{tot}=18.4\mu M$) by one or two orders of magnitude higher or lower. The base line is drawn in black. Lines in purple and red mean increasing the parameter, while lines in green and blue mean decreasing. In (D-H), similar procedures are performed on rates for monomer-dependent secondary nucleation, fragmentation, annealing and monomer conversion respectively. To be exact, parameters in (D) are taken according to those for the first subplot of Fig. 10 with $m_{tot}= 3.5\mu M$; parameters for (E) and (F) are the same as Fig. 9B with $m_{tot}= 11.5 \mu M$; (G) and (H) show a modified NE model by introducing an additional step of monomer conversion in (\ref{conversion}). Besides $k_c^+=0.01 s^(-1)$, $k_c^- = 0.001 s^(-1)$, $m_{tot}=18.4 \mu M$, other parameters are taken from Fig. 9A.}
		\label{Fig:mechanism}
\end{figure}

\subsection{How cytotoxicity arise}
After looking into so many fibrillation mechanisms, we still have no idea about the most important issue, \textit{i.e.} how the aggregation of misfolded amyloid proteins affects normal cell function and gives rise to amyloidosis? To answer this question, we first need to make sure which species of aggregates is responsible for cytotoxicity. For quite a long time, mature fibrils have been taken for granted as the major cause of cell damage \cite{chiti2006protein}. However, this view has been challenged again and again in recent years, with accumulated evidences coming from the
morphology, atomic structure and functions of oligomers and fibrils both in vitro and in vivo \cite{haass2007soluble,de2008insights}. Now heterogeneous
oligomers are generally believed far more toxic than mature fibrils \cite{walsh2007abeta, fandrich2012oligomeric, stefani2013oligomer}. Meanwhile, solvable monomers are considered as less harmful to cells even in a misfolded state \cite{small2001alzheimer, ono2009structure}.

In literature, several mechanisms have been proposed for the molecular basis of cytotoxicity caused by oligomeric species. Most arguments are based on the interaction between oligomers and lipid membrane. It is generally believed that the binding of oligomers could dramatically affect the shape and permeability of lipid membrane \cite{wong2009amyloid,williams2011membrane}, induce undesired pores or ion-channel-like structures \cite{capone2009amyloid,kawahara2010neurotoxicity}, and give rise to fatal abnormal ion leakage therefore \cite{engel2008membrane, xue2009fibril}. For instance, Demuro \textit{et al.} observed in Xenopus oocytes that A$\beta$(1-42) oligomers can lead to
abnormal Ca$^{2+}$ flux independent of ion-channel from 5 to 40 mer \cite{demuro2011single}. Schauerte \textit{et al.} further pointed out that hexamer is the smallest stable oligomer that can penetrate cell membrane, whereas 12 to 14-mers give rise to the largest ion current \cite{schauerte2010simultaneous}. By MD simulation, Jang \textit{et al.} proposed that 16- to 24-mers of A$\beta$ arrange into pore-like structures \cite{jang2009misfolded}, which are compatible with the pores observed by
atomic force microscopy \cite{quist2005amyloid, lin2001amyloid}. Meanwhile, Shafrir \textit{et al.} found that A$\beta$ pores is an assembly made up of six hexamers \cite{shafrir2010models}.

Alternative hypothesis based on cellular regulatory network suggests that in vivo the exposed flexible hydrophobic surfaces of oligomers
can promote aberrant protein interactions, deregulate cytosolic stress response \cite{bolognesi2010ans, olzscha2011amyloid}, trigger inflammatory responses, oxidative
damage \cite{querfurth2010mechanisms}, alter kinase and phosphatase activities, increase neurofibrillary tangles \cite{ono2009structure, demuro2005calcium, holtzman2011mapping}, change synaptic plasticity \cite{selkoe2011resolving, gong2003alzheimer, laferla2002calcium, turner2003roles} and so on.

Recently, Hong and his colleagues extended kinetic models for amyloid fibrillation and included the cell damage caused by oligomer formation too \cite{hong2015aKinetic}. Four basic assumptions have been put forward as a general criterion for modeling, \textit{i.e.} (1) the basic procedure of amyloid formation is well formulated by kinetic models; (2) cell damage is mainly caused by oligomers, rather than mature fibrils or monomers, through their binding to lipid membrane; (3) cytotoxicity is quantified through the amount of membrane-bounded oligomers (or leaked ion concentration in original manuscript); (4) oligomer binding does not affect the kinetics of amyloid formation. Or, in other words, the consumption of
oligomers during membrane binding could be neglected. The last assumption is generally unnecessary, however mathematically it offers a simple way to get rid of the feed-back influence of oligomer consumption during membrane binding and keep previous well-formulated equations of fibrillation kinetics unaffected.

Besides two equations for number concentration and mass concentration of aggregates $P(t)$ and $M(t)$, which characterize the fibrillation process under difference mechanisms in previous sections, the concentration of cells in different states, classified based on the number of membrane-bounded oligomers, evolves according to
\begin{eqnarray}
\frac{d}{dt}[C_j]=P_{oli}(t)\sum_{i=0}^{N-1}k_b^+(i)[C_i]-\sum_{i=1}^{N}k_b^-(i)[C_i],\quad j=0,\cdots,N\label{cell-damage}
\end{eqnarray}
where $[C_i]$ is the concentration of cells in state $i$, in which the lipid membrane has been bounded with $i$ oligomers. The current definition of cellular state is quite natural, since the condition of cell damage is positively correlated with the amount of bounded oligomers according to assumption (3) above. The total cell concentration $c_{tot}=\sum_{i=0}^{N}[C_i]$ is a constant, in which $N$ stands for the maximal number of oligomers allowed to be bounded to the same membrane. $k_b^+(i)$ and $k_b^-(i)$ are rate constants for oligomer binding and unbinding, which potentially depend on the cellular state. According to \cite{gregory2008quantitative,gregory2009magainin}, $k_b^+\sim(2-5)\times10^5M^{-1}s^{-1}$ and $k_b^-\sim5-300s^{-1}$ on average. $P_{oli}(t)$ is the number concentration of oligomers at time $t$ and can be expressed as a function of $M(t)$ and $P(t)$ according to the moment-closure method shown in section of how to coarse-grain model. However, it is still uncertain whether $M_{oli}$ (the mass concentration of oligomers) or even more complicated functions are required to include those toxic species and their relative damage to cells explicitly.

Especially, a simple two-state model including only normal and damaged cells is obtained when $N=1$, based on which the fraction of damaged cells is given by
\begin{eqnarray}
[C_1]/c_{tot}=\int_0^tk_b^+P_{oli}(\tau){\rm e}\bigg\{-\int_{\tau}^t[k_b^+P_{oli}(\sigma)-k_b^-]d\sigma\bigg\}d\tau.
\end{eqnarray}
Once the time course for $P_{oli}(t)$ is determined through the fibrillation kinetics, a full knowledge about the progress of cell damage caused by oligomers binding could be obtained. A similar approach by examining the leaked ion concentration has been adopt by Hong \textit{et al.} \cite{hong2015aKinetic} and applied to $\beta_2m$ and IAPP fibrils induced membrane leakage with great success (see Fig. 13).

\begin{figure}[h]
	\centering
		\includegraphics[width=0.8\textwidth]{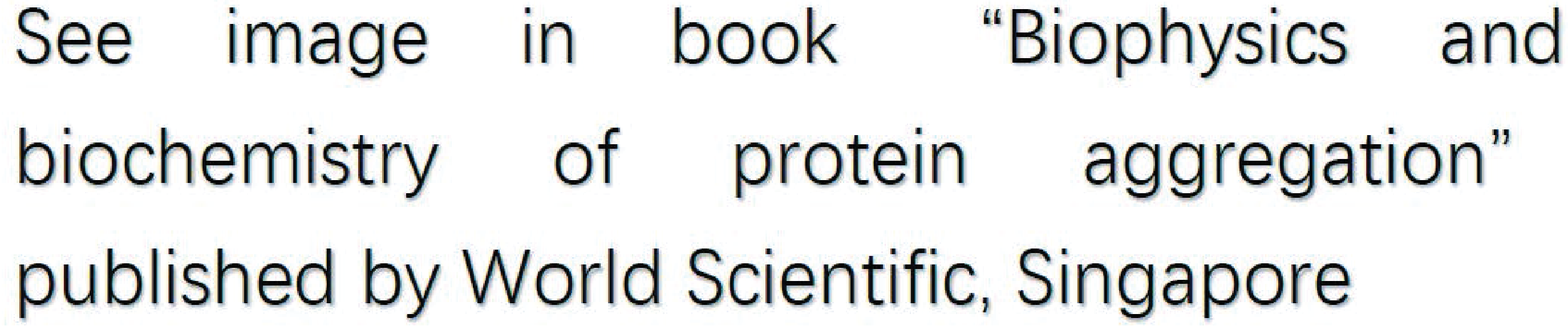}
		\caption{hIAPP fibril growth and
hIAPP-induced membrane leakage: (A) fibril mass
concentration measured by ThT fluorescence; (B)
the amount of membrane leakage under different
initial protein concentration; (C) the half-time for
fibril formation and membrane leakage; and (D)
the amount of membrane leakage under different
concentrations of seeds. Image is taken from \cite{hong2015aKinetic} with copyright permission, but all fittings are re-performed according to (\ref{cell-damage}) with $k_n^+=4\times10^{-5}M^{-1}s^{-1}, k_e^+=1.2\times10^{5}M^{-1}s^{-1}, k_e^-=0s^{-1}, k_f^+=7\times10^{-5}s^{-1}, k_f^-=1\times10^{4}M^{-1}s^{-1}, k_b^+=1.5\times10^{5}M^{-1}s^{-1}, k_b^-=2.5\times10^{-3}s^{-1}, n=1, n_c=2, n_o=40$. Particularly, in (D) $m_{tot}=2.5\times10^{-5}M, k_b^+=6\times10^{5}M^{-1}s^{-1}, k_b^-=1\times10^{-3}s^{-1}$ and $M(0)/P(0)=400$.}
		\label{Fig:NEF model}
\end{figure}

\subsection{How to manipulate fibrillation}
A central aim for examining various fibrillation mechanisms is to manipulate the kinetics of amyloidosis. Based on different targets, there are basically two large groups of approaches to achieve this goal. The indirect approach is to control the environment, in which amyloid fibrillation takes place. Here the word ``indirect'' means we will not manipulate amyloid proteins or fibrils directly. Alternatively, by varying the temperature, pH value, salt concentration \textit{etc.}, reaction rate constants for different processes will be changed accordingly. As a result, we can either accelerate or decelerate any given amyloid fibrillation, as well as increase or decrease the population of any oligomeric and fibrillar species. To be concrete, by decreasing the elongation rate, monomers will be overpopulated; otherwise, the population of mature fibrils will be enlarged. To increase the population of oligomers is a bit complex, which requires increasing the rate of primary nucleation and decreasing that of elongation at the same time.

According to the classical transition state theory (TST) for elementary chemical reactions, the rate constant for a given process is determined by the Arrhenius equation \cite{eyring1935activated, anslyn2006modern}
\begin{eqnarray}
k=A\exp[-E_{act}/(k_BT)],
\end{eqnarray}
where $A$ is referred to as the frequency factor, and $E$ is regarded as the activation energy, which in principle is a function of various experimental conditions, $E_{act}= E_{act}$(temperature, pH value, salt concentration \textit{etc.}). Therefore, if we know how the activation energy for various elementary fibrillation processes relies on conditions quantitatively, we can control the fibrillation kinetics as well as the population of each species freely as we wish.

Although for many cases the indirect approach is quite useful, it suffers some intrinsic limitations, \textit{e.g.} it is difficult to make a big change of some fibrillation conditions in experiments; some conditions may play a complex role in determining the reaction rate and no quantitative or qualitative description is available; in many cases, changing one condition may affect almost all processes at the same time, which makes it almost impossible to perform analysis.

The direct approach means to manipulate the concentration of each species directly. Varying the initial monomer concentration and seeds concentration are two most popular and effective methods in vitro, whose effects on fibrillation kinetics and species populations have been fully appreciated in previous sections. In vivo, directly adding or removing monomers and seeds becomes less promising. Therefore, specific binding through antibodies and chaperons is introduced to control the fibrillation kinetics as well as species population.

Antibodies and chaperons are both well-known for their specificity in binding to certain molecular structures. In principle, we can diminish any monomeric, oligomeric and fibrillar species by introducing proper antibody or chaperon into the system, which therefore allows us to manipulate each elementary fibrillation process independently. For example, nano-particles have been widely used in literature to inhibit amyloid fibrillization, induce fibril dissociation and mitigate neurotoxicity \cite{cabaleiro2008inhibition, yoo2011inhibition, liao2012negatively}. Similarly, many chaperon molecules, like Hsp70 family members, are known for their ability to inhibit and reverse the formation of amyloid aggregates \cite{huang2006heat, liberek2008chaperones, xu2013influence}. Now, searching for various promising antibodies and chaperons to inhibit amyloid fibrillation, or to prevent and cure amyloidosis as a potential goal, becomes very popular in this field (see Fig. 14 as an exmaple). Though in many cases, the molecular basis for why it works has not yet been clarified.

To mathematically probing various potential binding effects of antibodies (or chaperons) on amyloid fibrillation is a systematic and laborious task, and has not been developed into a mature theoretical framework. Here we just look into one particular example to show how to do it in principle. The mechanism of antibody blocking fibril ends and stopping the elongation process could be formulated into
\begin{eqnarray}
&&\frac{d}{dt}[B]=-k_b^+P_{free}[B]+k_b^-P_{bound},\label{binding1}\\
&&\frac{d}{dt}P_{free}=k_nm^{n_c} + k_2m^{n_2}M\underline{-k_b^+P_{free}[B]+k_b^-P_{bound}},\label{binding2}\\
&&\frac{d}{dt}M=k_e^+mP_{free}-k_e^-P_{free},\label{binding3}
\end{eqnarray}
where $P_{free}=P-P_{bound}$ and $P_{bound}=b_{tot}-[B]$ are concentrations of free fibril ends and antibody blocked fibril ends separately. $[B](t)$ is the free antibody concentration at time $t$, and $b_{tot}$ is the total concentration of antibodies, which is conserved during reactions. $k_b^+$ and $k_b^-$ are rate constants for antibody binding and unbinding respectively. In a similar way, antibody binding oligomers and blocking primary nucleation or secondary nucleation, chaperons binding fibrils and reversing amyloid formation, \textit{etc.} could be modeled and will not be addressed further.

In the presence of antibody or chaperon, model analysis becomes far more difficult. However, under one particular condition interested in experiments, \textit{i.e.} the rates for antibody  (or chaperon) binding and unbinding are much faster than the fibrillation speed, we can safely apply the Partial Equilibrium Approximation (PEA) to (\ref{binding1}) and suppose terms on the right-hand side cancelling each other at every moment, meaning $-k_b^+[B]P+k_b^-(b_{tot}-[B])=0$. Then class results for $P$ and $M$ without antibody binding are recovered. This leads to an important conclusion that, in order to manipulate amyloid fibrillation, the rate for antibody unbinding must be slower than that of fibrillation.

\begin{figure}[h]
	\centering
		\includegraphics[width=0.8\textwidth]{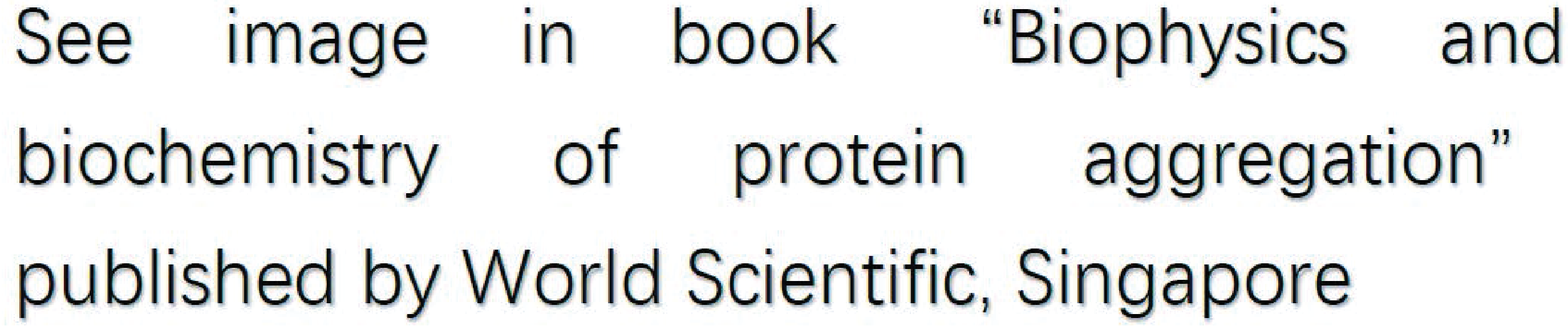}
		\caption{Brichos slows down A$\beta$42 aggregation by inhibiting the surface catalyzed secondary nucleation. (A)-(C): From left (blue) to right (green), 0\%, 10\%, 15\%, 35\%, 50\%, 75\% and 100\% A$\beta$42 monomer equivalents of Brichos have been added respectively. The concentration of monomeric A$\beta$42 is 3 $\mu$M. (D)-(F): The blue line corresponds to the situation in the absence of Brichos. The green dashed lines in (D)-(F) respectively show predictions for the cases in which primary nucleation, elongation and secondary nucleation are inhibited. Image is taken from \cite{cohen2015molecular} with copyright permission.}
		\label{Fig:mechanism}
\end{figure}

\subsection{Summary}
\begin{figure}[h]
	\centering
		\includegraphics[width=0.8\textwidth]{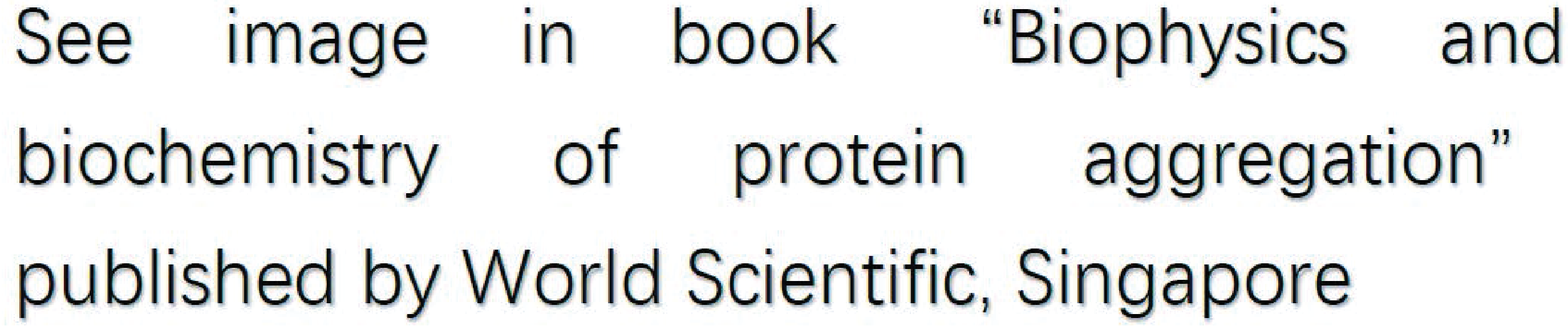}
		\caption{Scaling relationships between $k_{app}$, $t_{1/2}$, and $m_{tot}$ for eight fragmentation dominated amyloid proteins, \textit{i.e.} the yeast prion Sup35 NW region (purple triangles up), Csg $B_{trunc}$ (red squares), Ure2 protein (cyan pentacles), $\beta$2-microglobulin (brown stars), stefin B (blue cross), $\alpha$-synucleins (black triangles down), WW domain (yellow circles), and insulin (green diamonds). Image is taken from \cite{hong2011dissecting} with copyright permission.}
		\label{Fig:scaling}
\end{figure}

The fibrillation kinetics is a central issue in the study of amyloidosis and amyloid diseases. In the past several years, fruitful results and plenty of deep physical insights have been obtained in this direction. Among them, mathematical modeling based on chemical mass-action equations offers a unified framework to treat this problem. In the current section, we focus on how to apply kinetic modeling to explain various kinetic phenomena we have observed in real amyloid systems, which could be characterized and classified according to the half-time and fiber apparent growth rate as shown in Table I. Actually, the scaling relations between half-time of fibrillation (or fiber apparent growth rate) and monomer concentration provide us a useful way in practice to pick out the proper model, to classify amyloid systems based on their own fibrillation mechanisms (see Fig. \ref{Fig:scaling}) \textit{etc.}

\begin{table}[h]
    \begin{center}
    \begin{tabular}{|c||c|c|c|}
        \hline
        Mechanism & $t_{1/2}$ & $k_{app}$ & parameters\\
        \hline
        NE & $\lambda^{-1}$ & $\lambda$ & $\lambda=\sqrt{k_nk_e^+m_{tot}^{n_c}}$\\
        \hline
        NES & $\kappa_s^{-1}\ln(\kappa_s/\lambda)$ & $\kappa_s$ & $\kappa_s=\sqrt{k_e^+k_2m_{tot}^{n_2+1}}$\\
        \hline
        NEF & $\kappa_f^{-1}\ln(\kappa_f/\lambda)$ & $\kappa_f$ & $\kappa_f=\sqrt{k_e^+k_f^+m_{tot}}$\\
        \hline
        NE* & $\sqrt{1+\alpha}\lambda^{-1}$ & $\lambda/\sqrt{1+\alpha}$ & $\alpha=m_{tot}/K_m$\\
        \hline
        NE*S & $\sqrt{1+\alpha}\kappa_s^{-1}\ln(\kappa_s/\lambda)$ & $\kappa_s/\sqrt{1+\alpha}$ & \\
        \hline
        NE*F & $\sqrt{1+\alpha}\kappa_f^{-1}\ln(\kappa_f/\lambda)$ & $\kappa_f/\sqrt{1+\alpha}$ & \\
        \hline
        NES* & $\sqrt{1+\beta}\kappa_s^{-1}\ln(\kappa_s/\lambda)$ & $\kappa_s/\sqrt{1+\beta}$ & $\beta=(m_{tot}/K_s)^{n_2}$\\
        \hline
        NE*S* & $\sqrt{(1+\alpha)(1+\beta)}\kappa_s^{-1}\ln(\kappa_s/\lambda)$ & $\kappa_s/\sqrt{(1+\alpha)(1+\beta)}$ & \\
        \hline
        seeded E & $\nu^{-1}$ & $\nu$ & $\nu=k_e^+P_0$\\
        \hline
        seeded E* & $\alpha\nu^{-1}$ & $\nu/\alpha$ & \\
        \hline
    \end{tabular}
    \end{center}
    \caption{A summary of half-time and apparent fiber growth rate under different fibrillation mechanisms. Capital E stands for elongation, S for surface catalyzed secondary nucleation, F for fragmentation and $*$ for saturation of corresponding processes. ``Seeded'' means conditions with pre-added seeds. In general, fragmentation can be regarded as a special case of secondary nucleation with $n_2=0$.}
\end{table}

\section{Mathematical and application foundations}
The former section is wholly based on macroscopic kinetic models for the mass concentration and number concentration of fibrils, but why we can adopt this simple picture and how its accuracy is compared to models at a molecular lever have never been addressed. For this purpose, here we are going to the mathematical foundation of kinetic modeling, in which the mathematical linkage between models at the microscopic scale (molecular lever) and macroscopic scale (what we adopted in the former section) will be clarified in quantity. Interestingly, this linkage provides us ways to reconstruct the full fiber length distribution from a knowledge of mass concentration and number concentration of fibrils in a high accuracy, which is generally believed to be very difficult as an inverse problem. Finally, issues on how to convert model predictions into experimental observations, how to determine unknown model parameters and how to perform reliable global fittings are discussed too, in order to provide a relatively comprehensive review on the kinetic aspect.

\subsection{How to quantify fibrillation kinetics}
In order to quantify the kinetics of amyloid fibrillation, let's introduce the fiber length distribution $\{[A_i], i\geq1\}$, each of whose components represents the concentration of aggregates containing exactly $i$ protein molecules. Especially, $muiv[A_1]$ stands for the concentration of monomers. According to different reaction schemes for amyloid aggregation which we have addressed in details in the main text, the time evolution of $\{[A_i](t)\}$ will be characterized through a group of coupled ordinary differential equations (without taking the spacial distribution into consideration) by using laws of mass-action. This leads to the so-called ``microscopic chemical kinetic equations'', since they contain a full knowledge of the fiber length distribution.

On the other hand, either due to resolution limitation of experiments which makes it impossible to obtain a full spectrum of fiber length distribution, or for a purpose to speed up simulation without taking too many details into consideration, a simple coarse-grained description is preferred. Among various candidates, formulation involving two macroscopic quantities -- the number concentration and mass concentration of aggregates (including both oligomers and fibrils)
\begin{eqnarray}
P=\sum\limits_{i={n_c}}^{\infty} {[A_i]},\quad M=\sum\limits_{i={n_c}}^{\infty} {i\cdot[A_i]},
\end{eqnarray}
where $n_c$ stands for the critical nucleus size, is the most popular and welcomed. A basic reason is that this formulation actually provides a simplest self-consistent way to examine the amount of amyloid fibrils formed inside a given system, a quantity we are most interested in.

From above definition, we can see that $P$ and $M$ are actually the zeroth and first-order moments of fiber length distribution $\{[A_i]\}$ for $i\geq n_c$. The total mass concentration $m_{tot}=\sum\limits_{i=1}^{\infty} {i\cdot[A_i]}$ is another often used first-order moment. In particular, if oligomer concentration $[A_i]$ ($2\leq i\leq n_c-1]$) could be neglected, we will have $m_{tot}=m(t)+M(t)$, an equality representing the mass conservation of total protein molecules during fibrillation. In principle, other high order moments could also be introduced into the formulation in order to achieve a high accuracy, but in many cases it is not very worthwhile due to the sacrifice of both simplicity and efficiency. In literature, the governing equations of moments are referred to as ``macroscopic chemical kinetic equations'' in contrast to microscopic formulation based on fiber length distribution.

The relation between macroscopic and microscopic chemical kinetic equations is an interesting question. In fact, a whole branch of statistical physics is dealing with such problems on how to derive macroscopic quantities and their governing kinetic equations from a knowledge of microscopic descriptions \cite{landau1980statistical}. There are enormous investigations, fruitful results and endless debates in history which are far beyond the scope of this paper. But on current specific topic, almost all issues could be solved by some well-formulated moment-closure methods. We will come back to this point with all necessary details in the next section.

At last, we want to introduce the half time of fibrillation $t_{1/2}$ and the apparent fiber growth rate $k_{app}$ \cite{hong2011dissecting}, \textit{i.e.}
\begin{eqnarray}
M(t_{1/2})=\frac{M(0)+M(\infty)}{2},\quad k_{app}=\frac{1}{m_{tot}}\frac{dM}{dt}\bigg|_{t=t_{1/2}}.
\end{eqnarray}
These two quantities are essential to characterize the kinetics of amyloid fibrillation in an empirical way (see (\ref{ThT})), even without referring to any models or analytical solutions.
Two additional quantities appear in literature from time to time too. One is the lag time, which is defined as $t_{lag}=t_{1/2}-1/(2k_{app})$. The lag time has an intuitive physical meaning, which measures how long an amyloid system has to wait in order to accumulate enough seeds to pass the phase dominated by primary nucleation. In the presence of secondary nucleation, the lag time becomes prominent due to the sigmoid fibrillation profile, and could be greatly shortened by an introduction of initial seeds. There are also other ways to define the lag time in literature, \textit{e.g.} the time when the mass concentration of aggregates reaches $1\%$ of its static value \cite{librizzi2005kinetic}. Obviously, our definition is more natural and meaningful in physics. The other quantity is related to the speed of fibril growth, its maximal value to be exact. However, in most cases the maximal fiber growth rate $k_{max}$ relies on the whole fibrillation profile, which makes it difficult to determine and use.

\subsection{How to coarse-grain model}
In last section, we have introduced the fiber length distribution and its moments in different orders, like the number concentration and mass concentration of aggregates. We have also claimed there is a direct connection between the macroscopic and microscopic chemical kinetic equations. Now we are going to address this point based on the method of moment closure.

Without loss of generality, we consider the reaction scheme proposed for the length-dependent fragmentation \cite{hong2013simple}. According to laws of mass-action, the time evolution of fiber length distribution obeys following equations
\begin{eqnarray}
&&\frac{d[A_1]}{dt}=-n_ck_n^+[A_1]^{n_c}+n_ck_n^-[A_{n_c}]-k_e^+[A_1]\sum_{j=n_c}^{\infty}[A_j]+k_e^-\sum_{j=n_c+1}^{\infty}[A_j],\label{microeq-1}\\
&&\frac{d[A_i]}{dt}=k_e^+[A_1]([A_{i-1}]-[A_i])-k_e^-([A_{i}]-[A_{i+1}])+2\sum_{j=n_c+i}^{\infty}k_f^+(i,j-i)[A_j]\nonumber\\
&&-\sum_{j=n_c}^{i-n_c}k_f^+(j,i-j)[A_i]-2\sum_{j=n_c}^{\infty}k_f^-(i,j)[A_i][A_j]+\sum_{j=n_c}^{i-n_c}k_f^-(j,i-j)[A_j][A_{i-j}]\nonumber\\
&&+(k_n^+[A_1]^{n_c}-k_n^-[A_i]-k_e^+[A_1][A_{i-1}]+k_e^-[A_i])\delta_{i,n_c},\qquad i\geq n_c.\label{microeq-2}
\end{eqnarray}
As a consequence, it is straightforward to show that the number concentration $P(t)$ and mass concentration $M(t)$ of aggregates evolve according to
\begin{eqnarray}
&&\frac{d}{dt}P=k_n^+(m_{tot}-M)^{n_c}-k_n^-[A_{n_c}]+\sum_{i=n_c}^{\infty}\sum_{j=i+n_c}^{\infty}k_f^+(i,j-i)[A_j]-\sum_{i=n_c}^{\infty}\sum_{j=n_c}^{\infty}k_f^-(i,j)[A_i][A_j],\label{closure-P}\\
&&\frac{d}{dt}M=n_ck_n^+(m_{tot}-M)^{n_c}-n_ck_n^-[A_{n_c}]+k_e^+(m_{tot}-M)P-k_e^-P+k_e^-[A_{n_c}],\label{closure-M}
\end{eqnarray}
which clearly are not closed, as unknown variables $\{[A_i]\}$ have not been expressed through $P$ and $M$.

To solve this problem, a systematic moment closure method based on Maximum Entropy Principle \cite{jaynes1957information,jaynes1982rationale} has been proposed and applied with great success \cite{hong2013simple,tan2013modeling,hong2015aKinetic} (as shown in Fig. 16). Namely, we seek for solutions of following constrained optimization problem, \textit{i.e.}
\begin{eqnarray}
max &&S(\{[A_i]\})=-k_B\sum_{i=n_c}^{\infty}([A_i]\ln[A_i]-[A_i]),\\
s.t. &&\sum_{i=n_c}^{\infty}[A_i]=P,\; \sum_{i=n_c}^{\infty}i\cdot[A_i]=M,\;[A_1]+\sum_{i=n_c}^{\infty}i\cdot[A_i]=m_{tot}.
\end{eqnarray}
It could also be translated into an equivalent variational problem through the method of Lagrangian multiplier,
\begin{eqnarray}
\frac{\delta}{\delta[A_i]}\bigg[S(\{[A_i]\})/k_B+\lambda_1\bigg(\sum_{i=n_c}^{\infty}[A_i]-P\bigg)+\lambda_2\bigg(\sum_{i=n_c}^{\infty}i\cdot[A_i]-M\bigg)+\lambda_3\bigg([A_1]+\sum_{i=n_c}^{\infty}i\cdot[A_i]-m_{tot}\bigg)\bigg]=0,
\end{eqnarray}
where $\lambda_1$, $\lambda_2$ and $\lambda_3$ are Lagrangian multipliers. The solution of above equation is given by
\begin{eqnarray}
&&[A_1]={\rm e}(\lambda_3),\\
&&[A_i]={\rm e}[\lambda_1+i(\lambda_2+\lambda_3)],
\end{eqnarray}
based on which $\lambda_1$, $\lambda_2$ and $\lambda_3$ are related to $P(t)$, $M(t)$ and $m_{tot}$ as
\begin{eqnarray}
&&\lambda_1=\ln\bigg[\frac{P^2}{M-(n_c-1)P}\bigg]-n_c\ln\bigg[\frac{M-n_cP}{M-(n_c-1)P}\bigg],\\
&&\lambda_2=\ln\bigg[\frac{M-n_cP}{M-(n_c-1)P}\bigg]-\ln(m_{tot}-M),\\
&&\lambda_3=\ln(m_{tot}-M).
\end{eqnarray}
Put these formulas back into (\ref{closure-P}) and (\ref{closure-M}), we obtain desired macroscopic chemical kinetic equations solely concerning with $P(t)$ and $M(t)$.

In \cite{hong2013simple}, it has been proven that above moment closure method based on Maximum Entropy Principle is mathematically equivalent to Partial Equilibrium
Approximation on fiber elongation, which assumes the elongation process is much faster than primary nucleation and fragmentation. Therefore, for given
number concentration $P(t)$ and mass concentration $M(t)$ of aggregates, each component of fiber length distribution $\{[A_i]\}$ is considered in quasi-equilibrium with each other, which is exactly the way how we are able to express $[A_i](t)$ as a function of $P(t)$, $M(t)$ and $m_{tot}$.

\begin{figure}[h]
	\centering
		\includegraphics[width=0.6\textwidth]{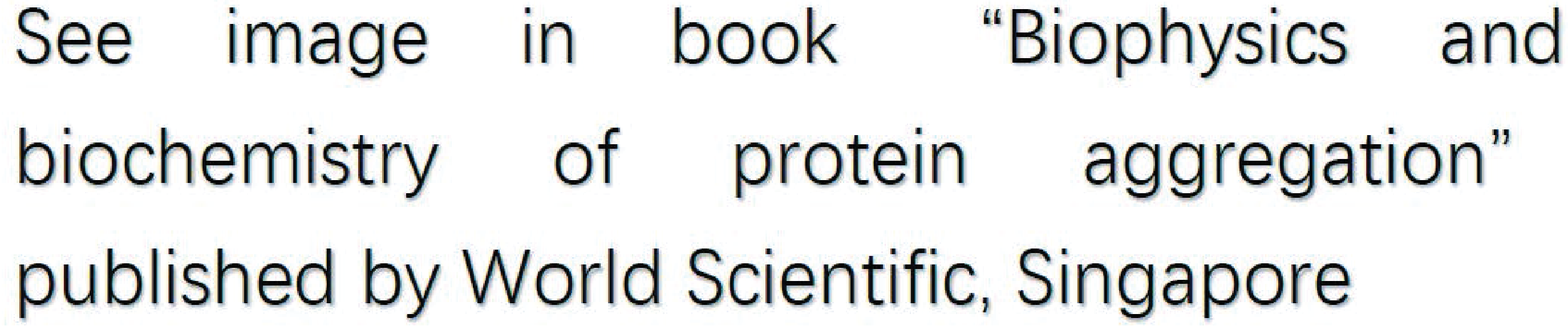}
		\caption{Accuracy of moment-closure method through comparisons on the fiber mass concentration, number concentration and fiber length distribution in (A-C). Values obtained from microscopic kinetic equations (\ref{microeq-1}) and (\ref{microeq-2}) are drawn in circles and that from moment-closure methods in solid lines. (D) Application to the polymerization of WW domain. Image is taken from \cite{hong2013simple}  with copyright permission.}
		\label{Fig:mechanism}
\end{figure}

\subsection{How to reconstruct fiber length distribution from moments}
In previous section, we have focused on how to simplify the model by coarse-graining, which turns out to be a very promising approach with tremendous successful applications. However, during coarse-graining, we are facing with an inevitable loss of information. Original knowledge of full fiber length distribution has been compressed into that about only two macroscopic moments -- the number and mass concentration of aggregates to be exact. Is it possible to reconstruct original fiber length distribution based on a knowledge of these two moments? This is a question not only of mathematical interest, but also with great practical usage. In principle, inverse problems are generally very difficult to solve and do not admit a unique solution \cite{fu2001characterization,poschel2003kinetics,tarantola2005inverse}. But, in current case without considering filaments fragmentation and annealing, as we are so lucky to have a complete understanding about underlying microscopic processes, the full fiber length distribution at any time could be explicitly extracted just from one moment -- mass concentration of aggregates as well as some knowledge about the initial length distribution. A brief derivation is listed as follows.

Without loss of generality, let us start with following microscopic model, including primary nucleation, elongation and secondary nucleation \cite{michaels2014asymptotic},
\begin{eqnarray}
&&\frac{d[A_i]}{dt}=k_e^+m(t)([A_{i-1}]-[A_i])+k_nm(t)^{n_c}\delta_{i,n_c}+k_2m(t)^{n_2}\big[m_{tot}-m(t)\big]\delta_{i,n_2},\quad i\geq n_c
\end{eqnarray}
where $n_2\geq n_c>0$. Introduce the generating function \cite{lando2003lectures}
\begin{eqnarray}
C(z,t)=\sum_{j=n_c}^{\infty}z^j\cdot[A_j](t),
\end{eqnarray}
which is a natural mathematical generalization of the physical moments. Especially, we have $P(t)=C(z=1,t)$ and $M(t)=\frac{\partial C(z,t)}{\partial z}|_{z=1}$. In fact, the ful fiber length distribution could be recovered from the generating function, \textit{i.e.} $[A_j](t)=\frac{1}{j!}\frac{\partial^jC(z,t)}{\partial z^j}\big|_{z=0}$ for $i\geq n_c$, which means the generation function is a one-to-one mapping of the fiber length distribution.

It is straightforward to show that the generating function satisfies following equation
\begin{eqnarray}
\frac{\partial C(z,t)}{dt}=k_e^+m(t)(z-1)C(z,t)+k_nm(t)^{n_c}z^{n_c}+k_2m(t)^{n_2}[m_{tot}-m(t)]z^{n_2}.
\end{eqnarray}
Define a new time scale $\tau(t)=\int_0^tk_e^+m(s)ds$, which acts as the characteristic time for fiber elongation, above equation could be rewritten as
\begin{eqnarray}
\frac{\partial C(z,\tau)}{d\tau}=(z-1)C(z,\tau)+\bigg\{k_nm(\tau)^{n_c}z^{n_c}+k_2m(\tau)^{n_2}[m_{tot}-m(\tau)]z^{n_2}\bigg\}\frac{\partial t}{\partial \tau},
\end{eqnarray}
whose solution is given by
\begin{eqnarray}
C(z,\tau(t))=\int_0^te^{-(z-1)[\tau(s)-\tau(t)]}\bigg\{k_nm(s)^{n_c}z^{n_c}+k_2m(s)^{n_2}[m_{tot}-m(s)]z^{n_2}\bigg\}ds+C(z,0)e^{(z-1)\tau(t)}.
\end{eqnarray}
Now the fiber length distribution at any given time $t$ could be calculated through following formula
\begin{eqnarray}
[A_j](t)=\frac{1}{j!}\frac{\partial^jC(z,t)}{\partial z^j}\bigg|_{z=0}&=&\int_0^t\Theta_{j-n_c}(t,s)k_nm(s)^{n_c}ds+\int_0^t\Theta_{j-n_2}(t,s)k_2m(s)^{n_2}[m_{tot}-m(s)]ds\nonumber\\
&&+\sum_{k=n_c}^{j}\Theta_{j-k}(t,0)[A_k](0),
\end{eqnarray}
where $\tau(t)=\int_0^tk_e^+m(s)ds$ and $\Theta_{k}(t,s)=e^{\tau(s)-\tau(t)}[\tau(t)-\tau(s)]^k/k!$. During the calculation, we use identities $0^0=1$ and $0!=1$.

Above formula provides the mathematical foundation on how to extract the full fiber length distribution at any time just based on the time course of monomer concentration (or mass concentration of aggregates $M(t)$) as well as the initial fiber length distribution. This is a quite astonishing result, as we have mentioned that inverse problems are usually extremely difficult to solve and do not admit a unique solution in general. Our success in this case could be contributed to two reasons: one is we have a complete knowledge on the microscopic kinetics which governs the time evolution of fiber length distribution; the other is both primary nucleation and secondary nucleation can solely affect the concentration of single species $[A_{n_c}]$ or $[A_{n_2}]$. The only way to perform a global change of the fiber length distribution is elongation, which follows a Poisson process characterized by the integral kernel $\Theta_{k}(t,s)$ with intrinsic time scale $\tau(t)=\int_0^tk_e^+m(s)ds$, since each monomer association is obviously random and independent of each other.

In the presence of filaments fragmentation and annealing, an additional global change of the fiber length distribution will be introduced besides elongation. How to include these two processes into above picture explicitly is an unsolved problem. In particular, Michaels \textit{et al.} derived approximate solutions for length-independent fragmentation in open and close systems \cite{michaels2015length}. While for general models with length-dependent fragmentation and annealing, an empirical method has been proposed, whose theoretical foundation lies on the fact that during the procedure of moment-closure, the microscopic fiber length distribution is expressed through macroscopic moments based on Maximum Entropy Principle. The thus obtained fiber length distribution directly determines the accuracy of macroscopic equations, which as a consequence could be adopted as an empirical candidate to approximate the exact fiber length distribution without introducing new errors. This empirical approach has been applied to examine a fragmentation-only model and matches perfectly with numerical solutions and experimental data for fiber-like PI264-b-PFS48 micelles under sonication as shown in Fig. 17 \cite{tan2013modeling}. In addition, similar procedure has been shown effective for a more complicated model, including primary nucleation, elongation and length-dependent fragmentation \cite{hong2012link}.

\begin{figure}[h]
	\centering
		\includegraphics[width=0.7\textwidth]{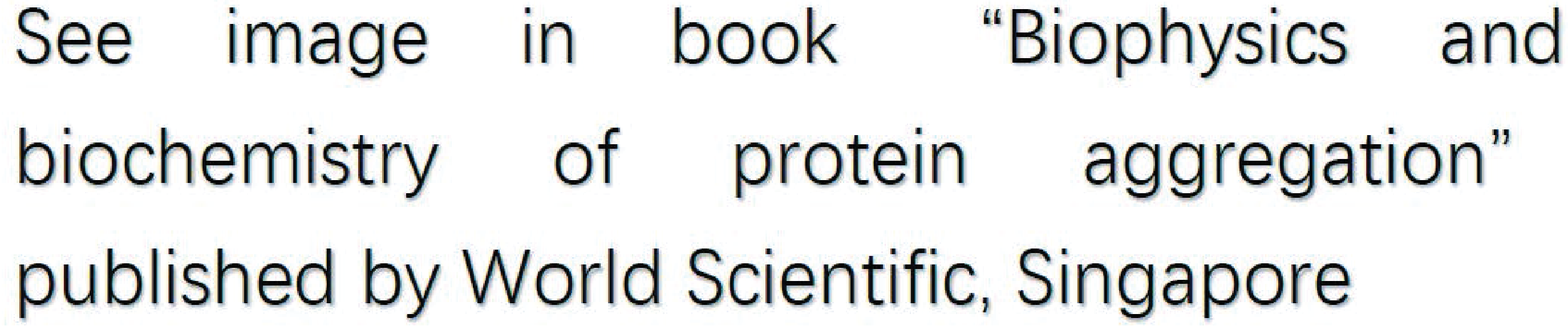}
		\caption{Accuracy of approximate fiber length distribution constructed from moment-closure method for a fragmentation-only model. Comparisons were made among TEM measurements of fiber-like PI264-b-PFS48 micelles under sonication (black symbols), exact fiber length distribution in (\ref{microeq-1}) and (\ref{microeq-2}) (red solid curves), and
approximate fiber length distributions (brown and blue dashed curves). Image is taken from \cite{tan2013modeling})  with copyright permission.}
		\label{Fig:mechanism}
\end{figure}

\subsection{How to make a connection to experiments}
According to Nilsson \cite{nilsson2004techniques}, there are three criteria that define a protein aggregate as an amyloid fibril: green birefringence upon staining with Congo Red, fibrillar morphology, and $\beta$-sheet secondary structure. Based on these three criteria, plenty of novel instruments and techniques have been developed to probe amyloid fibrils and their kinetics. Instead of providing a comprehensive review on each technique, like its basic setup, procedure and protocols, advantages and limitations, we will focus on the relation between those physical quantities we can measure and mathematical quantities we have defined for characterizing the fibrillation kinetics.

The most widely adopted technique to monitor the fibrillation kinetics is Thioflavin T (ThT) fluorescence. When it binds to $\beta$-sheet-rich structures, like those in amyloid aggregates, the dye displays enhanced fluorescence and a characteristic red shift of its emission spectrum \cite{khurana2005mechanism, biancalana2010molecular}. It has been shown that: (1) the ThT fluorescence intensity increases nearly linearly with the total amount of amyloid fibrils for several orders of magnitude; (2) the fluorescence intensity is independent of number concentration of amyloid fibrils if mass concentration is constant; (3) if the number concentration is fixed as in the extension kinetic study, the increase of average length of amyloid fibrils corresponds to an increase in the fluorescence intensity too \cite{naiki1989fluorometric, naiki1991kinetic}. These features make ThT fluorescence an unusually sensitive and efficient reporter for the mass concentration of aggregates $M(t)$, despite of limitations that ThT is not perfectly specific for amyloid fibrils, as well as some amyloid fibrils do not affect the fluorescence. In experiments, following empirical formula is often used to interpret the fluorescence intensity observed during amyloid fibrillation \cite{uversky2001metal},
\begin{eqnarray}
F(t)=F(0)+\frac{A}{1+{\rm e}[-k_{app}(t-t_{1/2})]},\label{ThT}
\end{eqnarray}
where $F(t)$ is the fluorescence intensity at time $t$, $F(0)$ is the background fluorescence intensity at the starting time, and $A$ is a fitting constant to correlate the mass concentration of aggregates with the absolute fluorescence intensity. Based on above formula, we can easily extract the apparent fiber growth rate $k_{app}$ and half-time of fiber formation $t_{1/2}$ from data fitting.

The goal of any absorption spectroscopy, \textit{e.g.} Fourier transform infrared (FTIR) spectroscopy, ultraviolet-visible (UV) spectroscopy, circular dichroism (CD) \textit{etc.}, is to detect the presence of certain molecular structure by measuring the light absorbed at each wavelength. It has been established that, in FTIR, a peak near $1645cm^{-1}$ indicates random coil, $1655cm^{-1}$ for $\alpha$-helix and $1620-1640cm^{-1}$ for $\beta$-sheet \cite{jackson1995use}; while in far-ultraviolet ($170-250nm$) circular dichroism, a pronounced double minimum at $208$ and $222nm$ indicates $\alpha$-helical structure, and a single minimum at $204nm$ or $217nm$ reflects random-coil or $\beta$-sheet structure \cite{pelton2000spectroscopic, greenfield2006using}. Thus we can make a quantitative correlation between the fraction of certain structure (say $\beta$-sheet as an indicator of amyloid fibrils) with absorbed light intensity at the corresponding wavelength. Here we take the CD spectroscopy as a simple example, in which the fraction of $\beta$-sheet structure could be expressed through the measured ellipticity as \cite{li2001natural}
\begin{eqnarray}
\frac{C_{beta}}{C_{beta}+C_{coil}}=\frac{\Theta_{217}^{obs}-\Theta_{217}^{0}}{\Theta_{217}^{max}-\Theta_{217}^{0}},
\end{eqnarray}
where $\Theta_{217}^{0}$ is the ellipticity at $217nm$ at the beginning of time; while $\Theta_{217}^{obs}$ and $\Theta_{217}^{max}$ represent observed and maximal ellipticity at $217nm$ during the whole measurement. $C_{beta}$ and $C_{coil}$ represent the concentration of $\beta$-sheet and random coil structures separately.

Instead of light absorption, dynamic light scattering (DLS) is a technique that can be used to determine the size distribution profile of small particles in suspension or polymers in solution \cite{berne2000dynamic}. Although light scattering by particles in solution depends on a number of factors, the most relevant one is the ratio of particle size
with respect to quantity $\lambda/[2\pi\sin(\theta/2)]$, where $\lambda$ is the wavelength of incident light and $\theta$ is the angle of detection \cite{aragon1976theory, pallitto2001mathematical}. As the average length of amyloid fibrils is expected to get larger during amyloid fibrillation, the
intensity of light scattered by fibrils will be directly proportional to the mass concentration of aggregates \cite{berne1974interpretation, powers2008mechanisms},
\begin{eqnarray}
I_{scat}(t)=\frac{Q\lambda M(t)}{4l\sin(\theta/2)},
\end{eqnarray}
where $l$ is the monomer size, $Q$ is a constant depending on the setup.

Currently, transmission electron microscopy (TEM), atomic-force microscopy (AFM) and scanning-force microscopy (SFM) are most advanced techniques with highest resolution. According to \cite{rigden1996macmillan}, the smallest distance that can be resolved with a TEM is approximately $0.2-0.5nm$ and, for STM, a typical resolution of several tenths of one nanometer can be achieved. In comparison, the diameter of typical filaments is around $3-4nm$ and $8-10nm$ for mature fibrils, while the length may vary from hundreds of nanometers to a few micrometers \cite{kowalewski1999situ,ward2000fractionation,khurana2003general}. Therefore, we may use those instruments to directly observe the growth, inhibition, propagation and adaptation of single fibril and even its breakage and branching in real time \cite{ban2003direct, ban2004direct, ban2006direct, andersen2009branching}. With high-resolution images in hand, the fiber length distribution could be extracted to certain accuracy, based on which both the number concentration and mass concentration of aggregates could be obtained. As a conclusion, advanced high-resolution microscopies could provide us most detailed information about amyloid fibrils, both their morphology and kinetics, though at an great expense of money and time.

\subsection{How to determine model parameters}
How to determine unknown parameters is a key step in model application. Although it is as important as modeling itself, details behind parameter fitting have seldom been clarified. A major obstacle is that usually most model parameters are empirical and hard to be precisely determined by either experiments or fundamental principles in nature. Their accuracy and validity heavily depends on the experience of modelers.

Here we face with the same problem. In current study, the parameters, we adopted for describing amyloid fibrillation, cell damage and antibody inhibition in the main text, can be roughly divided into four groups.

The initial protein concentration $m_{tot}$, mass concentration $M_0$ and number concentration $P_0$ of initial seeds, total cell concentration $c_{tot}$ and antibody concentration $b_{tot}$ are generally pre-specified in the setup. They are not adjustable during the fitting and can be classified into Group I.

Group II contains those application-insensitive parameters, like the critical nucleus size $n_c$, the monomer disassociation rate $k_e^-$ and filaments annealing rate $k_f^-$ in some cases. Since their values have a minor influence on the model performance, according to Occam's Razor, they can set to a default value (usually $n_c=2$ and $k_e^-=k_f^-=0$). In mathematics, sensitivity analysis allows a systematical determination of model parameters in this group.

Group III contains those parameters which can be determined either by experiments or by some fundamental principles. For example, as discussed in section of how fibrils grow, the fiber elongation rate $k_e^+$ for different amyloid proteins has been measured by plenty of techniques in recent years. We can directly take their values from the literature giveing the same amyloid protein under similar experimental conditions.

The last group includes those freely adjustable model parameters. In most cases, the primary nucleation rate $k_n$, critical nucleus size $n_2$ and rate $k_2$ for surface-catalyzed secondary nucleation, fiber fragmentation rate $k_f^+$ are essential for modeling the fibrillation kinetics, and oligomer binding and unbinding rates $k_b^+$ and $k_b^-$ for modeling cell damage. The first four parameters can be gotten by performing global fitting of amyloid formation, while the last two are determined through data on cytotoxicity (a sensitivity analysis of model parameters could be learned from Fig. 18). In this sense, there will be no free tunable parameter left.

\begin{figure}[h]
	\centering
		\includegraphics[width=0.9\textwidth]{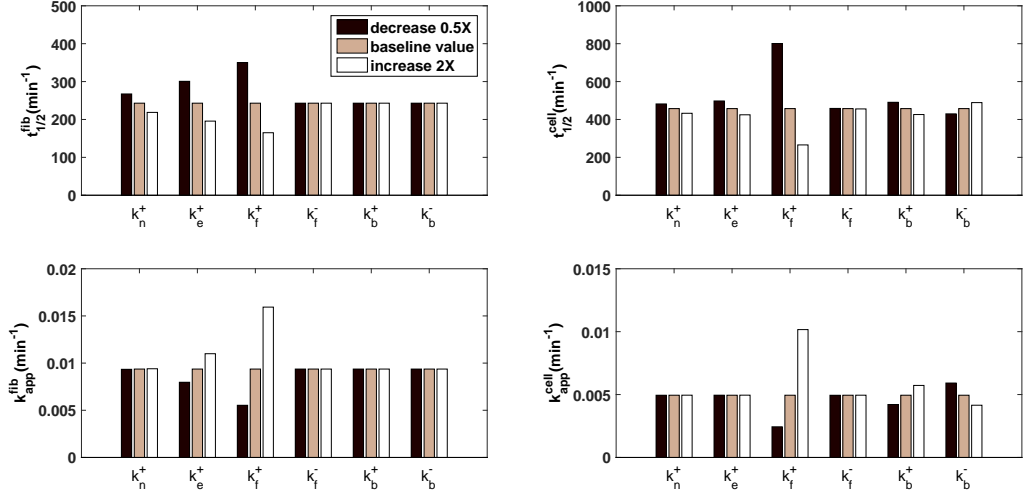}
		\caption{Model sensitivity on six adjustable parameters. The baseline values are taken
in accordance with those in Fig. 13B. All six reaction rates are changed by
either two times or one half with respect to their baseline values separately.}
		\label{Fig:mechanism}
\end{figure}

\subsection{How to perform global fitting}
As claimed by John von Neumann, ``With four parameters I can fit an elephant, and with five I can make him wiggle his trunk.'' An over-fit of experimental data with too many undetermined model parameters is encountered from time to time. Although a great effort has been dedicated to eliminate unnecessary model parameters as shown in last section, we are still facing with the problem how to perform a fitting reasonably and robustly. Global fitting, as suggested by its name, provides a nice way to partially solve this difficulty and make the fitting more reliable and promising.

The central idea of global fitting is to try to fit all data at the same time with the same parameters. In a such way, the redundancy in model parameters could be eliminated as much as possible and those key parameters will be highlighted. Although, the requirement of global fitting is quite natural from a theoretical view, in practice it is not so easy to perform it. One obstacle is data noise. In the presence of large noise, a global fit becomes impossible, which means we have to either try best to eliminate noise source and perform measurements as precisely as possible, or include the influence of noise into modeling at the beginning (\textit{e.g.} stochastic models). Another obstacle comes from the high dimensionality of the parameter space. Efficient global exploration methods, applicable to high dimensional space, has to be implemented during the procedure of fitting. In contrast, local exploration methods or those only valid for low dimensional space are not applicable. Finally, quantitative judgements or scores are needed in order to tell which group of model parameters gives the best fitting.

For this purpose, programs for global fitting have been developed for the kinetic models we have discussed in the main text \cite{meisl2016molecular}. Various global exploration methods, such as simulated annealing, genetic algorithm, particle swarm method \textit{ect.}, have been implemented into the program to avoid local trapping (private codes). Generally speaking, according to our own experience, simulated annealing is efficient and reliable in most cases; genetic algorithm consumes the longest CPU time among the three; while particle swarm is also quite slow, but in some cases it gives better results than simulated annealing. In these programs, nonlinear least-square regression is adopted to minimize
the sum of squared errors between experimental data and those predicted by the model. Now we plan to take the influence of noises on global fitting into consideration. Related works are going on.

\section{Conclusion}
In the past decades, due to the increasing interest on amyloid related diseases, a variety of amyloid proteins and their fibrillation processes have been investigated in details. In this self-contained reviewed, fruitful results on both thermodynamics and kinetics of amyloid fibrillation have been shown and discussed, with a purpose to provide a relatively comprehensive physical picture on what we know and what we do not know in this field. As a summary, facts we have learned from thermodynamic and kinetic modeling are:
\begin{enumerate}
      \item The existence of a critical fibrillar concentration below which fibrillar mass is negligible;
      \item Above the critical fibrillar concentration, the length distribution of the fibrils is exponential. This remains so even in the semi-dilute limit, if the fibril-fibril interactions are purely steric;
      \item If the fibrils exhibit attractive lateral interactions, there will be a strong tendency for phase separation of fibrils;
      \item Molecular mechanisms of amyloid fibrillation could be formulated into a group of microscopic chemical kinetic equations concerning with fiber length distribution;
      \item Low-order moments, a function of fiber length distribution, evolves according to macroscopic chemical mass-action equations derived by moment-closure method;
      \item Mass concentration and number concentration of aggregates are two most widely used moments and provide a well characterization of fibrils;
      \item In many cases, fiber length distribution could be reconstructed from moments exactly or approximately once the underlying fibrillation kinetics is known;
      \item Primary nucleation, elongation and secondary nucleation, including monomer-independent fragmentation and monomer-dependent surface catalyzed secondary nucleation, constitute a basic framework for amyloid fibrillation, which has been applied to many amyloid systems successfully;
      \item Conformational conversion of monomeric, oligomeric and fibrillar structures is crucial for amyloid fibrillation but easily neglected in kinetic modeling due to other rate-limiting steps;
      \item Effect of high monomer concentration on fiber elongation and surface catalyzed secondary nucleation could be explained by saturation;
      \item Oligomers binding to lipid membrane play a key role in cytotoxicity and can easily included in kinetic models;
      \item Quantitative knowledge on how to manipulate amyloid species and fibrillation kinetics could be learned from kinetic modeling, which provides us a power method to probe amyloidosis.
\end{enumerate}

In our opinion, answering the following questions would constitute fruitful research direction:
\begin{enumerate}
      \item How to incorporate various morphologies of amyloid fibrils into the thermodynamic picture?
      \item How to construct a complete picture of phase separation and transition for amyloid fibrillation?
      \item How to include oligomeric species explicitly in kinetic models?
      \item How to model lateral association of protofibrils or filaments into mature fibrils, like coiling and twisting?
      \item How to determine various reaction rate constants under a given fibrillation condition?
      \item How to systematically quantify the effect of antibody and chaperon in order to manipulate amyloid fibrillation?
      \item How to correlate amyloidosis with their fibrillation mechanisms at a molecular level?
      \item How to probe amyloid diseases with kinetic modeling?
\end{enumerate}

\section*{Acknowledgment}
This work was supported by the National Natural Science Foundation of China (Grants 11204150), Tsinghua University Initiative Scientific Research Program (Grants 20151080424) and the program of China Scholarships Council (CSC). Y.J.H also acknowledges the Postdoctoral Science Foundation of China (2015M581050).

\bibliographystyle{unsrt}
\bibliography{refs}

\end{document}